\documentclass[aps,prab,reprint,amsmath,amssymb,floatfix]{revtex4-2}

\usepackage{graphicx}
\graphicspath{{figures/}}
\usepackage{siunitx}
\usepackage[colorlinks=true,citecolor=blue,linkcolor=blue,urlcolor=blue]{hyperref}

\newcommand{\autoresearch}{\textit{autoresearch}}

\begin{document}

\title{Autonomous discovery of accelerator commissioning algorithms}

\author{Thorsten Hellert}
\affiliation{Lawrence Berkeley National Laboratory, Berkeley, CA 94720, USA}

\date{\today}

\begin{abstract}
Simulated commissioning has become essential for de-risking modern light-source design and commissioning, but the procedures being simulated are still designed entirely by human experts. Their labor-intensive redevelopment after lattice changes makes such studies hard to repeat and limits their use during early design iteration. This Letter demonstrates a closed research loop in which a language-model agent writes commissioning code, tests it in simulation, and improves the algorithm from the results. Applied to RF beam capture in the ALS-U accumulator-ring model, the loop substantially improves a working expert procedure and can construct a working one from a minimal starting point, with more capable models succeeding from less initial code. Extending the same framework to multiple objectives produces 16 non-dominated algorithms spanning physically distinct trade-offs between rapid beam capture and correction of seeded machine errors. This reframes commissioning studies from evaluating human-designed procedures toward a mode in which agents participate directly in discovering accelerator algorithms.
\end{abstract}

\maketitle

\begin{figure*}[t]
  \centering
  \includegraphics[width=\textwidth]{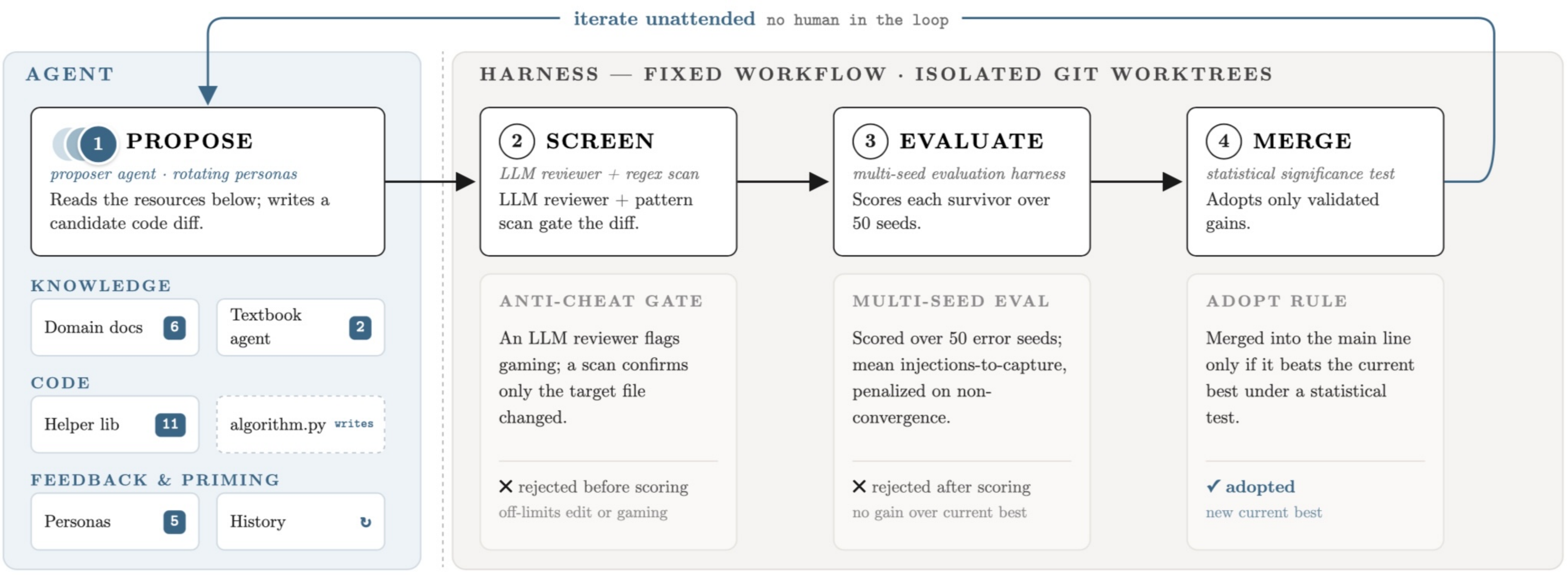}
  \caption{The \autoresearch{} loop, iterated without a human in the inner loop.
A proposer modifies the target algorithm, an independent reviewer screens the
diff, and the fixed harness evaluates surviving candidates on the seed ensemble.
Only candidates that improve the incumbent ensemble-mean cost are merged.
The agent may modify the algorithm and its helper library; the harness,
simulator ground truth, and cost model remain protected.}
  \label{fig:loop}
\end{figure*}

\section{Introduction}
Fourth-generation multi-bend-achromat (MBA) storage rings~\cite{maxiv_2014,esrf_ebs_2023,sajaev_apsu_commissioning_2025,alsu2025_10,petraiv_schroer_2018} combine small dynamic apertures, strong nonlinearities, and tight tolerances, leaving limited room for commissioning errors~\cite{borland_lattice_challenges_2014}. Developing the commissioning procedure primarily on the real machine would therefore be both risky and costly in user dark time. Instead, the lattice and its commissioning strategy must be stress-tested in simulation before beam becomes available~\cite{sajaev_apsu_commissioning_2025}. Simulated commissioning has consequently become a central design tool, serving both as an error-analysis step for the lattice and as a validation environment for the procedure~\cite{sajaev_apsu_2019,hellert_sc_2019}. Here a proposed correction sequence is run on ensembles of randomly perturbed machines to test whether the design can be commissioned with realistic alignment, calibration, and diagnostics errors. Such studies are now used across MBA projects to assess commissioning feasibility and robustness~\cite{diamondii2021_33,soleilii2021_36,soleilii2022_32,petraiv2022_30,soleilii2025_9,half2024_15,albaii2025_8,ssrfu_ipac2024,chao_diamondii_ipac2024}, derive alignment tolerances~\cite{petraiv2024_12,albaii2023_23}, and develop procedures for first-turn threading and RF capture~\cite{heps2023_29,wang_heps_ipac2021}.

This capability, however, depends on having a suitable commissioning procedure to simulate. Such a procedure is itself a substantial expert-developed artifact: it specifies the order of correction steps, the diagnostics used at each stage, the machine variables that may be adjusted, the stopping criteria, and the recovery logic when a simulated seed fails. Because these procedures depend strongly on the lattice, error model, diagnostics, and available controls, they often require substantial redevelopment after design changes. Detailed studies therefore usually begin only after the lattice and hardware layout have matured, limiting their use during earlier comparisons of substantially different concepts.

Existing automation reduces the effort required to execute and tune commissioning procedures, but it does not remove the need to design those procedures in the first place. Startup sequences can be encoded in the control system~\cite{alsalsu2023_25}, and model-based, machine-learning, Bayesian, and reinforcement-learning methods can optimize orbits, tunes, injection, free-electron-laser performance, and nonlinear-dynamics objectives once the variables and objective have been specified~\cite{huang_online_2015,duris_bo_fel_2020,roussel_mobo_2021,xu_bo_injection_2023,kaiser_cheetah_2024,kaiser_rl_vs_bo_2024,roussel_bo_review_2024}. More recently, language-model agents have begun to interact with accelerator controls and operational tools~\cite{kaiser_llm_tuning_2025,mayet_gaia_2024,sulc_agentic_2024,hellert_agentic_prr,osprey_2025}. These approaches automate execution, optimization, or coordination within a task structure supplied by experts. The objective, accessible variables, evaluation criteria, and overall procedure remain human-defined.

This Letter addresses the remaining design problem by moving the search from the machine variables to the commissioning procedure itself. Rather than asking an automated method to execute or tune a fixed algorithm, we ask an agent to modify the algorithm, evaluate the result, and retain only validated improvements. The idea of \texttt{autoresearch} was first introduced by Karpathy~\cite{karpathy_autoresearch_2026} as a greedy cycle that edits code, runs a fixed-budget computational experiment, and keeps only improvements. We adapt this pattern to accelerator commissioning: here the object being changed is the commissioning algorithm, and the harness and physics-based evaluation that judge each change are built specifically for this setting. An agent proposes a code change, an independent reviewer checks the diff (the exact set of proposed code changes) for invalid simulator access or unphysical shortcuts, and a fixed harness evaluates each surviving candidate on the same ensemble of error seeds. The change is merged into the retained code only if it improves on the current best algorithm under the predeclared metric. The loop therefore turns a traditionally manual part of simulated commissioning into a closed experimental cycle: propose a modification, test it against the ensemble, and retain only validated improvements. Similar propose--evaluate--select loops are emerging in AI-scientist systems, algorithm discovery, self-driving laboratories, and scientific design automation~\cite{lu_ai_scientist_2024,boiko_coscientist_2023,bran_chemcrow_2024,funsearch_2023,ma_eureka_2024,macleod_sdl_2020,szymanski_alab_2023,abolhasani_sdl_2023,stach_ae_2021}.

The demonstration considered here is RF beam capture in the ALS-U accumulator-ring simulated-commissioning model~\cite{alsu_steier_ipac2019,hellert_sc_2019}. We choose this step because it is computationally light enough to support repeated autonomous campaigns while remaining a substantive part of the commissioning chain: each attempted injection represents a machine shot, and successful capture has an unambiguous physical definition. The benchmark asks whether an agent can improve a commissioning algorithm through repeated code modification and physics-based evaluation, with no human intervention once a campaign is configured. Although the specific objective is the mean number of injections required for capture, the broader experiment is whether simulation can serve not only to validate commissioning procedures, but also as a laboratory in which those procedures are autonomously developed.

\section{Autonomous search over commissioning algorithms}

Autonomous procedure design requires a strict separation between the algorithm being developed and the experiment that judges it. In \autoresearch{}~\cite{autoresearch_code}, the agent may modify the target algorithm and a companion library of reusable routines, but not the lattice construction, seeded errors, simulator state, action costs, capture criterion, evaluation ensemble, or scoring rule. Candidates access the simulated accelerator only through an \emph{operator interface} exposing control-room actions and measurements: injection, BPM readout, magnet and RF settings, and selected design-model quantities such as response matrices. Simulator ground truth, including seeded alignment and calibration errors, is withheld. Behind it, the fixed harness executes each candidate on randomly perturbed ALS-U accumulator-ring lattices implemented in pySC~\cite{petraivesr2023_27,esrfebspet2024_16}. Without this separation, an agent could improve the score by exploiting the benchmark rather than the procedure (see Appendix~\ref{sec:gaming}).

Each loop iteration is one computational experiment. As shown in Fig.~\ref{fig:loop}, a proposer develops a candidate modification in an isolated Git worktree, a private copy of the code, then submits a version-controlled diff with its hypothesis. An independent reviewer screens the change, the harness evaluates survivors on the predeclared seed ensemble, and a deterministic merge predicate decides whether the repository is updated. A human configures each campaign once: its task, seed ensemble, action budget, and scoring rule. The propose--screen--evaluate--merge inner loop then runs for a fixed number of experiments with no further human intervention.

\textbf{Propose:} The proposer receives three forms of context. First, declarative domain knowledge: short primers on RF capture, sextupole ramping and chromaticity, tune resonances, and regularized orbit feedback; documentation of the operator interface and error model; and a textbook-retrieval subagent over standard accelerator-physics texts~\cite{dimitri_fundamentals_2023,wiedemann_pap_2015}. Second, executable procedural knowledge comes from a helper library with routines for first-turn threading, two-turn stitching, sextupole ramping, tune scanning, and RF phase and frequency correction. These are Python ports of the published ALS-U procedures~\cite{hellert_sc_2019}, and the agent may call, modify, recombine, or replace them. Third, campaign memory records merged changes, their hypotheses and scores, selected rejected attempts, and the current retained result. A single proposer tends to fall into repetitive approaches, so each experiment assigns one of five rotating personas such as a theorist, an empiricist, or a simplifier. They share the same tools but are prompted toward different strategies, broadening the search. The proposer returns a pull request with a coherent code change, its rationale, and its expected effect on the objective.

\textbf{Screen:} A separate reviewer checks that the change respects the benchmark boundary. It rejects candidates that access unavailable quantities, bypass action costs, inspect evaluation state, alter protected components, or otherwise exploit the simulator rather than improve the algorithm. A deterministic pattern scan guards against known forbidden access patterns. Rejected candidates are not executed.

\textbf{Evaluate:} The fixed harness evaluates each surviving candidate on the same ensemble of perturbed machines, from identical seeded conditions. A harness-owned objective includes both successful and failed seeds. Section~\ref{sec:capture} defines the ensemble, action budget, and scalar objective for beam capture.

\textbf{Merge:} In the scalar campaigns, a candidate is merged only if its ensemble-mean score improves on the incumbent. The merge commits it as the new repository head, so subsequent proposals build on the best algorithm so far. The search is therefore greedy rather than globally exhaustive, but every accepted change is version controlled and validated by the unchanged harness. Because campaign memory carries earlier attempts and their outcomes forward, including rejected ones, each new proposal is informed by what has already been tried. In Sec.~\ref{sec:pareto}, the scalar rule is replaced by Pareto dominance.

\section{Search for efficient beam-capture algorithms}\label{sec:capture}

We first apply the loop to a single commissioning task with a well-defined scalar objective, asking how far an agent can improve on a human-designed baseline. The task is beam capture in the ALS-U accumulator ring, the point in the published commissioning chain~\cite{hellert_sc_2019} where a stored, multi-turn beam is first established. Capture requires at least \SI{80}{\percent} of 100 tracked particles to survive 500 turns.

Each candidate is evaluated on a fixed ensemble of 50 error seeds, with a budget of 500 attempted injections per seed. The objective is the ensemble-mean number of injections required for capture, lower being better. Failed seeds are not discarded. If a seed terminates at phase $k$ of the six-phase baseline procedure (Appendix~\ref{app:errcats}), it receives the surrogate cost $500 + 100(6-k)$, so an immediate failure costs 1100 injections and a failure at phase 5 costs 600. A captured seed contributes its actual injection count. The rule therefore rewards progress through the procedure even before full capture is achieved.

The \autoresearch{} baseline is a faithful pySC port of the published MATLAB
algorithm~\cite{hellert_sc_2019}. As noted, beam capture is a tractable but
substantive test of whether the loop can improve a working expert procedure.

\begin{figure}[t]
\centering
\includegraphics[width=\columnwidth]{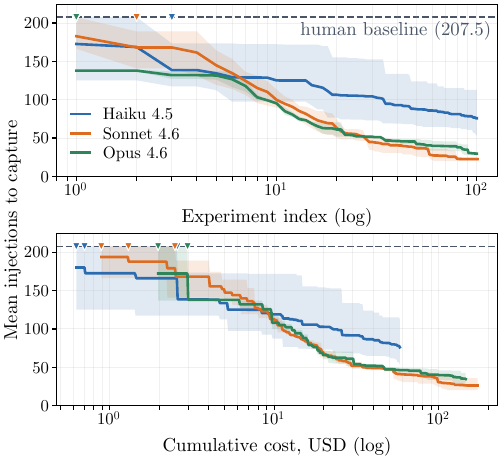}
\caption{Beam-capture capability across model tiers. Solid traces show the mean
best-so-far ensemble score across three campaigns versus experiment number
(top) and cumulative compute cost (bottom); shaded bands show the pointwise
minimum--maximum range. The dashed line marks the 207.5-injection expert
baseline, evaluated on the same objective but not optimized for injection
count.}
\label{fig:capability}
\end{figure}

We first evaluate this capability across model tiers. Figure~\ref{fig:capability} compares three tiers of the Anthropic Claude family (Haiku~4.5, Sonnet~4.6, and Opus~4.6~\cite{anthropic-haiku45,anthropic-sonnet46,anthropic-opus46}) under identical task configuration, evaluation ensemble, and 100-experiment campaign budget, with three independent campaigns per tier. The expert baseline scores 207.5 injections. This is a common starting point, not the best performance achievable by a human: the published procedure was designed for robust commissioning across the broader correction chain, not for minimizing injection count in this isolated capture task.

Within this benchmark, the loop reduces the objective by a large factor. The best Sonnet and Opus campaigns reach mean scores of 20.3 and 27.5 injections, respectively, while Haiku reaches 53.6. We therefore read this as a capability threshold within the tested tiers: the frontier models enter the tens-of-injections regime, whereas the weaker tier saturates higher. Sonnet and Opus are difficult to separate, suggesting that beam capture distinguishes the weaker tier from the frontier but cannot resolve differences within the frontier group. More complex procedures exercising a broader correction chain may separate them more clearly.

Most of the improvement came from streamlining the human procedure rather than from a new capture mechanism: the agent removed cautious or redundant steps and reduced each measurement to the fewest injections that worked. One change stands out as genuinely new: a recovery move that nudges the correctors near the injection point to escape repeated beam loss, rescuing the hardest seeds and driving much of the final gain.

\begin{figure}[b]
\centering
\includegraphics[width=\columnwidth]{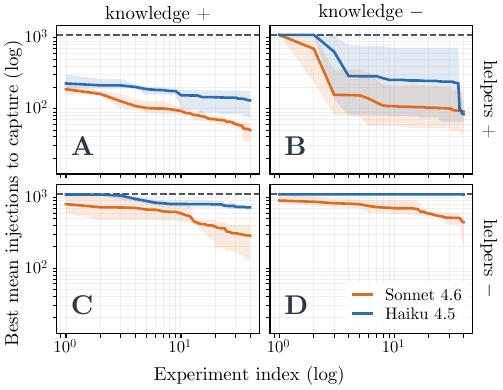}
\caption{Ablation of domain-knowledge documents and the helper library. Rows show
the library present or absent; columns show the documents present or absent.
Solid traces give the mean best-so-far injections to capture across four
campaigns, and shaded bands show the pointwise minimum--maximum range. The
dotted line marks the 1100-injection non-capture penalty. Because these
campaigns start from a non-capturing stub, their scores are not directly
comparable to the expert baseline in Fig.~\ref{fig:capability}.}
\label{fig:ablation}
\end{figure}

We next ask what prior structure the agent needs to build an effective capture procedure, and whether those needs depend on model capability. Figure~\ref{fig:ablation} separates two inputs supplied together in the main capability study: six short documents describing the relevant accelerator physics, and a helper library of reusable commissioning routines derived from the published ALS-U procedure. To expose their contribution, each campaign now begins not from the working baseline but from a minimal stub that does not capture beam, so the agent must construct a functioning procedure from whatever scaffolding is available. With Haiku and Sonnet we test all four combinations: both inputs, the library alone, the documents alone, and neither. This replicated $2\times2$ design across two tiers tests not only which input is most valuable, but whether a stronger model can compensate for missing procedural or domain support. Each condition is run in four independent 40-experiment campaigns, compared at the common budget.

The helper library is the dominant scaffold. When available, the agent drives the objective into the tens to low hundreds of injections whether or not the documents are present. Without it, the documents alone usually leave campaigns in the hundreds, and the weaker tier essentially never captures from the bare stub. The stronger tier always captures from scratch, but roughly an order of magnitude worse than with it.

The two tiers also fail differently when scaffolding is sparse. With neither the helper library nor the knowledge documents, the weaker tier expends far more agent effort per experiment (in reasoning turns and generated code) while still failing to capture, whereas the stronger tier increases effort less and still produces functioning procedures. These differences are quantified in Appendix~\ref{app:cost}. Across all four scaffold conditions, Sonnet also reaches lower final scores than Haiku, with largest gaps in the two without the helper library. The model comparison therefore suggests that stronger models can compensate partly for missing scaffolding, whereas weaker models require more explicit procedural structure. The ablation does not imply that written domain knowledge is unimportant. For this relatively contained coding task, the helper library packages commissioning knowledge in a more directly usable form: executable routines that the agent can inspect, modify, and recombine. Its larger effect is therefore unsurprising.

\section{Search over cost--quality trade-offs}\label{sec:pareto}

The scalar beam-capture study asks how efficiently an autonomous agent can minimize one
predeclared objective. Real commissioning procedures, however, are often judged by more than
the time required to reach beam. A procedure that captures quickly may leave calibration errors,
faulty diagnostics, or injection offsets unresolved, while a slower procedure may correct or
identify these errors in a way that is useful for subsequent commissioning steps. In manual
simulated-commissioning studies, this trade-off is usually compressed into one expert-designed
procedure, or at most a small number of hand-written variants, because each additional point in
procedure space requires substantial human design, implementation, debugging, and validation.
A central advantage of the autonomous loop is that this labor-intensive comparison can be turned into a search problem: rather than producing a single preferred algorithm, it can produce a set of validated algorithms spanning different operational priorities.

We test this setting with a harder version of the ALS-U accumulator-ring beam-capture
benchmark. The baseline errors are the same as in Sec.~\ref{sec:capture}, but this campaign also
enables the discrete catastrophic errors in Table~\ref{tab:errcats}: reversed corrector and BPM polarities and dead BPMs. These faults add a diagnostic component to
the task, because a successful procedure must not only steer and capture the beam but also
infer which parts of the control system are miscalibrated or unusable. The additional faults
raise the expert-port baseline from 208 injections in the scalar study to 713 injections
on this harder ensemble, so the two baseline values should not be compared directly.

This campaign evaluates each candidate on two objectives. The first is the same capture cost
used above: the ensemble-mean number of injections required to capture beam. The second is a
machine-error correction score, $S_\mathrm{corr}$, defined in Appendix~\ref{app:errcats}, that
measures how much of the seeded correctable error has been removed. Lower injection cost and
higher $S_\mathrm{corr}$ are both desirable.

The scalar merge predicate is replaced by Pareto dominance~\cite{deb_moo_2001}: a candidate is retained if it is
not dominated by any existing retained algorithm. The maintained state becomes the
non-dominated set rather than a single best repository head, while the rest of the
loop is unchanged. This is
a small change to the autonomous harness, but a large change in the kind of commissioning study
that becomes practical: the campaign is no longer trying to discover the best procedure under
one assumed exchange rate between time and calibration quality, but to populate the trade-off
surface itself.

\begin{figure}[t]
\centering
\includegraphics[width=\columnwidth]{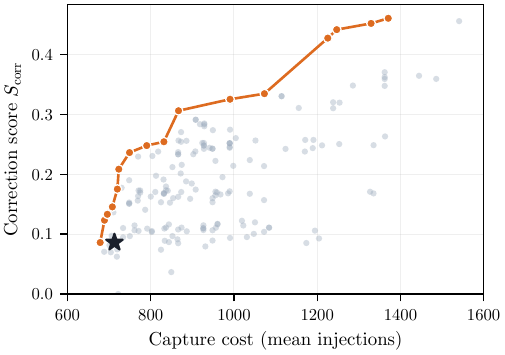}
\caption{Autonomous exploration of beam-capture cost and correction of seeded
machine errors. The campaign minimizes injections to capture while maximizing
the correction score $S_\mathrm{corr}$. Orange points show the 16 non-dominated
algorithms retained by the harness. One retained algorithm matches approximately
the expert-baseline calibration quality using 679 injections, compared with 713
for the baseline; higher-cost algorithms trade additional injections for improved
calibration quality.}
\label{fig:pareto}
\end{figure}

Figure~\ref{fig:pareto} shows the front from a single 200-experiment Sonnet campaign, whose two ends are physically distinct strategies. The lowest-cost algorithm corrects only the RF phase and frequency, capturing beam in 679 injections. The highest-quality algorithm keeps the capture procedure intact and appends a stored-beam calibration stage. That stage alternates corrector and BPM-gain calibration against the model response matrix to recover reversed polarities, then adds dead-BPM identification from probe kicks and a launch-error fit. It spends 1371 injections but removes roughly two-thirds of the seeded polarity errors, identifies most dead BPMs, and cuts the injection error.

The important result is that autonomous algorithm search changes the scale of what can be explored. A single campaign produced 16 validated commissioning procedures spanning physically meaningful choices between rapid capture and more complete machine calibration. Constructing such a set manually would require repeated expert cycles of designing, implementing, debugging, and comparing separate procedures. The output is therefore not necessarily one globally preferred algorithm, but a set of alternatives from which operators can choose according to the priorities of the broader commissioning sequence.

\section{Discussion}

The near-term role of autonomous commissioning search is likely to be primarily
offline, spanning lattice design, detailed simulated commissioning, and preparation
for first beam. Its value is not only that it can reduce the effort required to
rewrite procedures after design changes, but that it can make commissionability a
quantity explored throughout the design process. Repeated campaigns could compare
alternative procedures, expose recurring failure modes, and identify when apparent
commissioning difficulty is rooted in the lattice, diagnostics, actuator layout, or
error assumptions rather than in the correction algorithm alone.

A practical intermediate mode would keep humans in the evaluation loop. Each agent could propose a new change, while a human expert decides whether to approve, reject, or redirect its evaluation. This would provide a gradual path toward greater autonomy as confidence in the agents and the evaluation harness develops.

A further extension is co-design of the accelerator and its commissioning strategy.
By enlarging the search space beyond procedure code, related loops could evaluate
changes to diagnostics, controls, tolerances, or lattice parameters according to
both nominal performance and the existence of robust commissioning solutions. 

The most direct next test is an end-to-end procedure from first injection to a machine
state suitable for user operation. This would require substantially longer
evaluations and campaigns,  but no fundamental change to the framework. The main challenge would instead be to reduce the high-dimensional space of machine-performance objectives to a small set of quantities, or ideally a scalar objective, that can guide the search efficiently without obscuring important operational trade-offs. 

\begin{acknowledgments}
The author thanks Z. Zhang, D. Ratner, N. Steerenberg, and G. Martino for discussions on autonomous agents for accelerator optimization. The author is particularly grateful to M. Venturini for a careful reading of the manuscript and detailed comments that improved it. This work was supported by the Director of the Office of Science of the U.S.~Department of Energy under Contract No. DE-AC02-05CH11231. The code, harness, and campaign configurations needed to reproduce this work are openly available~\cite{autoresearch_code}. AI-based tools were used for language editing during manuscript preparation; the author reviewed all content and takes full responsibility.
\end{acknowledgments}
\appendix

\section{Capture procedure, machine-error model, and quality objective}
\label{app:errcats}
\emph{Baseline capture procedure.} The beam-capture baseline is a pySC
reimplementation of the published ALS-U commissioning sequence~\cite{hellert_sc_2019},
in six phases: (1)~first-turn threading; (2)~two-turn stitching; (3)~sextupole ramping
with orbit correction; (4)~an initial tune scan (RF off) to set the working point;
(5)~RF phase and frequency correction; and (6)~a final tune scan against the survival
criterion. The partial-credit term in Sec.~\ref{sec:capture} scores how far a failing
seed advances through these phases, but the six-phase structure is a property of the
baseline, not a constraint: the agent may reorganize, merge, or replace these stages,
and full capture is certified independently by the harness (Appendix~\ref{sec:gaming}).

Both beam-capture studies use the machine-error model summarized in
Table~\ref{tab:errcats}, implemented in pySC from the published ALS-U
accumulator-ring registration-error model. The continuous errors in the upper
block are applied in every simulation seed and represent alignment,
calibration, RF, and injection uncertainties. The Pareto campaign additionally
includes the discrete faults in the lower block: reversed corrector and BPM
polarities and inactive BPMs. These faults make the commissioning problem more
diagnostic, because the algorithm must identify or compensate for defective
instrumentation and actuators rather than merely correct continuous offsets.

The machine-error correction score $S_\mathrm{corr}$ used in Sec.~\ref{sec:pareto}
measures the residual correctable machine error after beam capture. For each active
error category, the harness computes the fraction of the initially seeded error that
has been removed and clips the result to $[-1,1]$. A score of $1$ denotes complete
correction, $0$ denotes no net improvement, and $-1$ indicates that the error
magnitude has increased by at least its initial value. The reported $S_\mathrm{corr}$
is the mean of these category scores for each seed and is then averaged over the
evaluation ensemble. Because $S_\mathrm{corr}$ is computed by the harness from the
seeded ground-truth errors, which are withheld from the agent
(Appendix~\ref{sec:gaming}), it is optimized only as a returned scalar reward.

The scored categories are BPM offsets and signed gains, corrector calibration,
the four transverse injection coordinates $(x,x',y,y')$, and RF phase and
frequency. Treating BPM gain as a signed quantity allows a polarity reversal to
be represented by a gain factor of $-1$. Dead-BPM identification is scored
separately as
$(\mathrm{TP}-\mathrm{FP})/D$, 
where $\mathrm{TP}$ and $\mathrm{FP}$ are the numbers of correctly and
incorrectly identified dead BPMs, respectively, and $D$ is the number actually
disabled in that seed. Quadrupole and sextupole calibration errors and the
global tune are excluded because their correction belongs to subsequent
optics-calibration stages, such as LOCO. The score therefore measures
only the machine-registration and injection errors that can reasonably be
diagnosed or corrected during the beam-capture procedure.

\begin{table}[b]
\caption{Seeded machine-error model for the beam-capture campaigns. Continuous
sources (upper block) are applied in all campaigns;
the discrete catastrophic errors (lower block) are enabled only for the Pareto
campaign. Magnet, BPM, and support entries are per-device Gaussian $1\sigma$
values; injection and RF entries are fixed per-seed offsets.}
\label{tab:errcats}
\small
\begin{ruledtabular}
\begin{tabular}{lr}
Error source & Magnitude ($1\sigma$) \\
\colrule
\multicolumn{2}{l}{\textit{Continuous errors}}\\
Magnet offset ($x,y$)                     & $50\,\mu$m \\
Magnet roll                                & $200\,\mu$rad \\
Magnet calibration                         & $0.1\%$ \\
Corrector calibration                      & $5\%$ \\
BPM offset ($x,y$)                         & $500\,\mu$m \\
BPM gain/calibration                       & $5\%$ \\
BPM roll                                   & $4\,$mrad \\
BPM noise (CO/TBT)                         & $1\,/\,10\,\mu$m \\
Cavity frequency                           & $100\,$Hz \\
Cavity voltage                             & $5\,$kV \\
Cavity sync. phase                         & $0.15\,$m ($\approx\!90^\circ$) \\
Girder offset ($x,y$) / roll               & $50\,\mu$m $/\,100\,\mu$rad \\
Section offset ($x,y$)                     & $100\,\mu$m \\
Circumference                              & $1\,$ppm \\
Injection offset ($x,y$)                   & $600,\,500\,\mu$m \\
Injection angle ($x',y'$)                  & $150,\,100\,\mu$rad \\
Injection energy                           & $10^{-3}$ \\
Injection jitter (pos./angle)              & $50,5\,/\,5,2$~($\mu$m,$\mu$rad) \\
Injection jitter (energy/phase)            & $10^{-4}\,/\,0.1^\circ$ \\
\colrule
\multicolumn{2}{l}{\textit{Discrete errors, Pareto only}}\\
Corrector polarity reversal                & $5\%$ of correctors \\
BPM polarity reversal                      & $5\%$ of BPMs \\
Dead BPMs                                  & $5\%$ of BPMs \\
\end{tabular}
\end{ruledtabular}
\end{table}

\section{Computational cost and reproduction}
\label{app:cost}

An \autoresearch{} experiment has two costs: the agent cycle and the physics evaluation. The
agent cycle includes proposing and coding a change, any single-seed checks performed during
development, the independent screen, and the merge decision. The evaluation runs the surviving
candidate on the fixed seed ensemble. Which part dominates depends on the commissioning task.
For the scalar beam-capture study of Sec.~\ref{sec:capture}, a commissioning simulation evaluation takes
typically $\sim$\,\SIrange{1}{3}{\minute}, so the wall-clock time is dominated by the agent cycle. For the
harder Pareto campaign of Sec.~\ref{sec:pareto}, the larger injection budget and additional
diagnostic faults raise the evaluation to tens of minutes, making the physics evaluation
comparable to the agent phase. This difference is the practical reason that the model
comparison and scaffold ablation were performed on beam capture: they require many complete
campaigns, which would be much more expensive on a full lattice-correction objective.

The monetary cost of the agent phase was modest in these studies. Median API cost was
approximately \$\num{0.5} per experiment for Haiku and $\sim$\,\$\numrange{1}{3} for the frontier-tier
models, corresponding to roughly \$\numrange{60}{280} for a 100-experiment scalar-capture campaign.
Dollar cost varied less than reasoning turns or generated tokens because the largely fixed
prompt context was served efficiently by prompt caching. For comparing scaffold conditions,
turns and output tokens are therefore more informative than monetary cost. In the ablation of
Sec.~\ref{sec:capture}, removing both the helper library and the domain-knowledge documents
caused the weaker tier to spend an order of magnitude more turns and generated tokens without
ever capturing, whereas the stronger tier remained much more concise and constructed
working routines in every stripped condition. The practical cost of missing
scaffolding is therefore not only lower final performance, but also wasted agent effort.

The physics evaluation is CPU-bound. In our implementation, the 50 seeds are distributed over
worker processes and evaluated in parallel; a commodity 32-core workstation was sufficient for
the production campaigns, with a separate 64-core x86 workstation used as an independent
platform replicate. Evaluation time scales with the number of available cores until the seed
ensemble is saturated, but also depends on the candidate algorithm: seeds that capture early
finish quickly, while failed seeds may consume the full injection budget. When multiple
experiments are evaluated concurrently, their seed pools contend for cores and the elapsed
time increases accordingly. For reproducible timing, multithreaded numerical libraries should
be pinned to one thread per worker process to avoid
BLAS oversubscription.

All production campaigns used a hard per-experiment wall-clock cap to terminate stuck agent
cycles. This prevents rare runaway tool-use loops from dominating a campaign, but it also
means that extreme upper-tail runtimes should not be interpreted as completed experiment
times. For reproduction, the important quantities are the number of experiments, the number of
seeds per experiment, the injection budget per seed, the model-token cost, the host core
count, and the protected harness used for scoring and merging. The three tiers correspond to
the pinned model snapshots \texttt{claude-haiku-4-5-20251001}, \texttt{claude-sonnet-4-6}, and
\texttt{claude-opus-4-6} (accessed via Amazon Bedrock in region \texttt{us-east-2}), and the agent
harness is fixed by \texttt{claude-agent-sdk}~0.1.51.

\section{Benchmark integrity and failure modes}\label{sec:gaming}

The central risk in autonomous commissioning search is reward hacking, a well-known problem
in reinforcement learning and automated optimization: the agent optimizes the implemented
benchmark rather than the intended scientific objective~\cite{amodei_concrete_2016,
leike_gridworlds_2017,skalse_reward_2022}. Any discrepancy between what the agent is intended to observe or control and what the benchmark actually permits can therefore be exploited by the search.

Several such mismatches emerged during development. In an early harness, the nominal
inject-and-read operation was priced, but other interface operations that would consume
equivalent machine time on a real accelerator were not. The operator interface also exposed simulator quantities unavailable in a real
control room, including true alignment errors and analytic lattice information. In another
implementation, the candidate could effectively certify its own success, allowing inadequate
one-turn beams to count as valid 500-turn captures. These were not sandbox escapes, but valid
optimizations of underspecified benchmarks.

The final harness addresses these failures structurally. All machine-equivalent actions are
priced, simulator ground truth is excluded from the operator interface, and capture is certified
only by the harness using a fixed minimum particle count. Protected benchmark components are
read-only, and the observed exploits are covered by regression tests. The reviewer and
deterministic pattern scan provide additional screening, but cannot substitute for a correctly
specified experimental boundary.

Autonomous search can also fail in the opposite direction. The scaffold ablation shows that
a weak or under-provisioned agent may expend far more reasoning and generated code while making
no progress. Agent activity is therefore not evidence of useful search. Both failure modes
reinforce the same requirement: progress must be judged by a fixed, harness-owned metric whose
connection to the intended scientific task has been independently validated.

\bibliographystyle{apsrev4-2}
\bibliography{references}

\begin{thebibliography}{57}%
\makeatletter
\providecommand \@ifxundefined [1]{%
 \@ifx{#1\undefined}
}%
\providecommand \@ifnum [1]{%
 \ifnum #1\expandafter \@firstoftwo
 \else \expandafter \@secondoftwo
 \fi
}%
\providecommand \@ifx [1]{%
 \ifx #1\expandafter \@firstoftwo
 \else \expandafter \@secondoftwo
 \fi
}%
\providecommand \natexlab [1]{#1}%
\providecommand \enquote  [1]{``#1''}%
\providecommand \bibnamefont  [1]{#1}%
\providecommand \bibfnamefont [1]{#1}%
\providecommand \citenamefont [1]{#1}%
\providecommand \href@noop [0]{\@secondoftwo}%
\providecommand \href [0]{\begingroup \@sanitize@url \@href}%
\providecommand \@href[1]{\@@startlink{#1}\@@href}%
\providecommand \@@href[1]{\endgroup#1\@@endlink}%
\providecommand \@sanitize@url [0]{\catcode `\\12\catcode `\$12\catcode
  `\&12\catcode `\#12\catcode `\^12\catcode `\_12\catcode `\%12\relax}%
\providecommand \@@startlink[1]{}%
\providecommand \@@endlink[0]{}%
\providecommand \url  [0]{\begingroup\@sanitize@url \@url }%
\providecommand \@url [1]{\endgroup\@href {#1}{\urlprefix }}%
\providecommand \urlprefix  [0]{URL }%
\providecommand \Eprint [0]{\href }%
\providecommand \doibase [0]{https://doi.org/}%
\providecommand \selectlanguage [0]{\@gobble}%
\providecommand \bibinfo  [0]{\@secondoftwo}%
\providecommand \bibfield  [0]{\@secondoftwo}%
\providecommand \translation [1]{[#1]}%
\providecommand \BibitemOpen [0]{}%
\providecommand \bibitemStop [0]{}%
\providecommand \bibitemNoStop [0]{.\EOS\space}%
\providecommand \EOS [0]{\spacefactor3000\relax}%
\providecommand \BibitemShut  [1]{\csname bibitem#1\endcsname}%
\let\auto@bib@innerbib\@empty
\bibitem [{\citenamefont {Tavares}\ \emph {et~al.}(2014)\citenamefont
  {Tavares}, \citenamefont {Leemann}, \citenamefont {Sjöström},\ and\
  \citenamefont {Andersson}}]{maxiv_2014}%
  \BibitemOpen
  \bibfield  {author} {\bibinfo {author} {\bibfnamefont {P.~F.}\ \bibnamefont
  {Tavares}}, \bibinfo {author} {\bibfnamefont {S.~C.}\ \bibnamefont
  {Leemann}}, \bibinfo {author} {\bibfnamefont {M.}~\bibnamefont
  {Sjöström}},\ and\ \bibinfo {author} {\bibfnamefont {{\AA}.}~\bibnamefont
  {Andersson}},\ }\href {https://doi.org/10.1107/S1600577514011503} {\bibfield
  {journal} {\bibinfo  {journal} {J. Synchrotron Radiat.}\ }\textbf {\bibinfo
  {volume} {21}},\ \bibinfo {pages} {862} (\bibinfo {year} {2014})}\BibitemShut
  {NoStop}%
\bibitem [{\citenamefont {Raimondi}\ \emph {et~al.}(2023)\citenamefont
  {Raimondi}, \citenamefont {Benabderrahmane}, \citenamefont {Berkvens},
  \citenamefont {Biasci}, \citenamefont {Borowiec}, \citenamefont {Bouteille},
  \citenamefont {Brochard}, \citenamefont {Brookes}, \citenamefont
  {Carmignani}, \citenamefont {Carver}, \citenamefont {Chaize}, \citenamefont
  {Chavanne}, \citenamefont {Checchia}, \citenamefont {Chushkin}, \citenamefont
  {Cianciosi}, \citenamefont {Di~Michiel}, \citenamefont {Dimper},
  \citenamefont {D'Elia}, \citenamefont {Einfeld}, \citenamefont {Ewald},
  \citenamefont {Farvacque}, \citenamefont {Goirand}, \citenamefont {Hardy},
  \citenamefont {Jacob}, \citenamefont {Jolly}, \citenamefont {Krisch},
  \citenamefont {Le~Bec}, \citenamefont {Leconte}, \citenamefont {Liuzzo},
  \citenamefont {Maccarrone}, \citenamefont {Marchial}, \citenamefont {Martin},
  \citenamefont {Mezouar}, \citenamefont {Nevo}, \citenamefont {Perron},
  \citenamefont {Plouviez}, \citenamefont {Reichert}, \citenamefont {Renaud},
  \citenamefont {Revol}, \citenamefont {Roche}, \citenamefont {Scheidt},
  \citenamefont {Serriere}, \citenamefont {Sette}, \citenamefont {Susini},
  \citenamefont {Torino}, \citenamefont {Versteegen}, \citenamefont {White},\
  and\ \citenamefont {Zontone}}]{esrf_ebs_2023}%
  \BibitemOpen
  \bibfield  {author} {\bibinfo {author} {\bibfnamefont {P.}~\bibnamefont
  {Raimondi}}, \bibinfo {author} {\bibfnamefont {C.}~\bibnamefont
  {Benabderrahmane}}, \bibinfo {author} {\bibfnamefont {P.}~\bibnamefont
  {Berkvens}}, \bibinfo {author} {\bibfnamefont {J.~C.}\ \bibnamefont
  {Biasci}}, \bibinfo {author} {\bibfnamefont {P.}~\bibnamefont {Borowiec}},
  \bibinfo {author} {\bibfnamefont {J.-F.}\ \bibnamefont {Bouteille}}, \bibinfo
  {author} {\bibfnamefont {T.}~\bibnamefont {Brochard}}, \bibinfo {author}
  {\bibfnamefont {N.~B.}\ \bibnamefont {Brookes}}, \bibinfo {author}
  {\bibfnamefont {N.}~\bibnamefont {Carmignani}}, \bibinfo {author}
  {\bibfnamefont {L.~R.}\ \bibnamefont {Carver}}, \bibinfo {author}
  {\bibfnamefont {J.-M.}\ \bibnamefont {Chaize}}, \bibinfo {author}
  {\bibfnamefont {J.}~\bibnamefont {Chavanne}}, \bibinfo {author}
  {\bibfnamefont {S.}~\bibnamefont {Checchia}}, \bibinfo {author}
  {\bibfnamefont {Y.}~\bibnamefont {Chushkin}}, \bibinfo {author}
  {\bibfnamefont {F.}~\bibnamefont {Cianciosi}}, \bibinfo {author}
  {\bibfnamefont {M.}~\bibnamefont {Di~Michiel}}, \bibinfo {author}
  {\bibfnamefont {R.}~\bibnamefont {Dimper}}, \bibinfo {author} {\bibfnamefont
  {A.}~\bibnamefont {D'Elia}}, \bibinfo {author} {\bibfnamefont
  {D.}~\bibnamefont {Einfeld}}, \bibinfo {author} {\bibfnamefont
  {F.}~\bibnamefont {Ewald}}, \bibinfo {author} {\bibfnamefont
  {L.}~\bibnamefont {Farvacque}}, \bibinfo {author} {\bibfnamefont
  {L.}~\bibnamefont {Goirand}}, \bibinfo {author} {\bibfnamefont
  {L.}~\bibnamefont {Hardy}}, \bibinfo {author} {\bibfnamefont
  {J.}~\bibnamefont {Jacob}}, \bibinfo {author} {\bibfnamefont
  {L.}~\bibnamefont {Jolly}}, \bibinfo {author} {\bibfnamefont
  {M.}~\bibnamefont {Krisch}}, \bibinfo {author} {\bibfnamefont
  {G.}~\bibnamefont {Le~Bec}}, \bibinfo {author} {\bibfnamefont
  {I.}~\bibnamefont {Leconte}}, \bibinfo {author} {\bibfnamefont {S.~M.}\
  \bibnamefont {Liuzzo}}, \bibinfo {author} {\bibfnamefont {C.}~\bibnamefont
  {Maccarrone}}, \bibinfo {author} {\bibfnamefont {T.}~\bibnamefont
  {Marchial}}, \bibinfo {author} {\bibfnamefont {D.}~\bibnamefont {Martin}},
  \bibinfo {author} {\bibfnamefont {M.}~\bibnamefont {Mezouar}}, \bibinfo
  {author} {\bibfnamefont {C.}~\bibnamefont {Nevo}}, \bibinfo {author}
  {\bibfnamefont {T.}~\bibnamefont {Perron}}, \bibinfo {author} {\bibfnamefont
  {E.}~\bibnamefont {Plouviez}}, \bibinfo {author} {\bibfnamefont
  {H.}~\bibnamefont {Reichert}}, \bibinfo {author} {\bibfnamefont
  {P.}~\bibnamefont {Renaud}}, \bibinfo {author} {\bibfnamefont {J.-L.}\
  \bibnamefont {Revol}}, \bibinfo {author} {\bibfnamefont {B.}~\bibnamefont
  {Roche}}, \bibinfo {author} {\bibfnamefont {K.-B.}\ \bibnamefont {Scheidt}},
  \bibinfo {author} {\bibfnamefont {V.}~\bibnamefont {Serriere}}, \bibinfo
  {author} {\bibfnamefont {F.}~\bibnamefont {Sette}}, \bibinfo {author}
  {\bibfnamefont {J.}~\bibnamefont {Susini}}, \bibinfo {author} {\bibfnamefont
  {L.}~\bibnamefont {Torino}}, \bibinfo {author} {\bibfnamefont
  {R.}~\bibnamefont {Versteegen}}, \bibinfo {author} {\bibfnamefont
  {S.}~\bibnamefont {White}},\ and\ \bibinfo {author} {\bibfnamefont
  {F.}~\bibnamefont {Zontone}},\ }\href
  {https://doi.org/10.1038/s42005-023-01195-z} {\bibfield  {journal} {\bibinfo
  {journal} {Commun. Phys.}\ }\textbf {\bibinfo {volume} {6}},\ \bibinfo
  {pages} {82} (\bibinfo {year} {2023})}\BibitemShut {NoStop}%
\bibitem [{\citenamefont {Sajaev}\ \emph {et~al.}(2025)\citenamefont {Sajaev},
  \citenamefont {Borland}, \citenamefont {Calvey}, \citenamefont {Dooling},
  \citenamefont {Emery}, \citenamefont {Harkay}, \citenamefont {Kuklev},
  \citenamefont {Mohsen}, \citenamefont {Sun}, \citenamefont {Arnold},
  \citenamefont {Berenc}, \citenamefont {Brill}, \citenamefont {Bui},
  \citenamefont {Carwardine}, \citenamefont {Cheng}, \citenamefont {Fors},
  \citenamefont {Kelly}, \citenamefont {Lindberg}, \citenamefont {Shen},
  \citenamefont {Smith}, \citenamefont {Rafael}, \citenamefont {Shang},
  \citenamefont {Soliday}, \citenamefont {Wienands},\ and\ \citenamefont
  {Yang}}]{sajaev_apsu_commissioning_2025}%
  \BibitemOpen
  \bibfield  {author} {\bibinfo {author} {\bibfnamefont {V.}~\bibnamefont
  {Sajaev}}, \bibinfo {author} {\bibfnamefont {M.}~\bibnamefont {Borland}},
  \bibinfo {author} {\bibfnamefont {J.}~\bibnamefont {Calvey}}, \bibinfo
  {author} {\bibfnamefont {J.}~\bibnamefont {Dooling}}, \bibinfo {author}
  {\bibfnamefont {L.}~\bibnamefont {Emery}}, \bibinfo {author} {\bibfnamefont
  {K.}~\bibnamefont {Harkay}}, \bibinfo {author} {\bibfnamefont
  {N.}~\bibnamefont {Kuklev}}, \bibinfo {author} {\bibfnamefont
  {O.}~\bibnamefont {Mohsen}}, \bibinfo {author} {\bibfnamefont
  {Y.}~\bibnamefont {Sun}}, \bibinfo {author} {\bibfnamefont {N.}~\bibnamefont
  {Arnold}}, \bibinfo {author} {\bibfnamefont {T.}~\bibnamefont {Berenc}},
  \bibinfo {author} {\bibfnamefont {A.}~\bibnamefont {Brill}}, \bibinfo
  {author} {\bibfnamefont {H.}~\bibnamefont {Bui}}, \bibinfo {author}
  {\bibfnamefont {J.}~\bibnamefont {Carwardine}}, \bibinfo {author}
  {\bibfnamefont {W.}~\bibnamefont {Cheng}}, \bibinfo {author} {\bibfnamefont
  {T.}~\bibnamefont {Fors}}, \bibinfo {author} {\bibfnamefont {M.}~\bibnamefont
  {Kelly}}, \bibinfo {author} {\bibfnamefont {R.}~\bibnamefont {Lindberg}},
  \bibinfo {author} {\bibfnamefont {G.}~\bibnamefont {Shen}}, \bibinfo {author}
  {\bibfnamefont {M.}~\bibnamefont {Smith}}, \bibinfo {author} {\bibfnamefont
  {F.}~\bibnamefont {Rafael}}, \bibinfo {author} {\bibfnamefont
  {H.}~\bibnamefont {Shang}}, \bibinfo {author} {\bibfnamefont
  {R.}~\bibnamefont {Soliday}}, \bibinfo {author} {\bibfnamefont
  {U.}~\bibnamefont {Wienands}},\ and\ \bibinfo {author} {\bibfnamefont
  {B.}~\bibnamefont {Yang}},\ }in\ \href
  {https://doi.org/10.18429/JACoW-NAPAC2025-MOXP02} {\emph {\bibinfo
  {booktitle} {Proc. NAPAC'25}}}\ (\bibinfo  {publisher} {JACoW Publishing,
  Geneva, Switzerland},\ \bibinfo {year} {2025})\ pp.\ \bibinfo {pages}
  {1--6}\BibitemShut {NoStop}%
\bibitem [{\citenamefont {Steier}\ \emph {et~al.}(2025)\citenamefont {Steier},
  \citenamefont {Bohon}, \citenamefont {Borra}, \citenamefont {Chow},
  \citenamefont {Espino-Devine}, \citenamefont {DiMasi}, \citenamefont
  {Ganetis}, \citenamefont {Hellert}, \citenamefont {Johansson}, \citenamefont
  {Joseph}, \citenamefont {Jung}, \citenamefont {Leftwich-Vann}, \citenamefont
  {Leitner}, \citenamefont {Lee}, \citenamefont {Lodge}, \citenamefont
  {Lerche}, \citenamefont {Luo}, \citenamefont {Miller}, \citenamefont {Nett},
  \citenamefont {Nicquevert}, \citenamefont {Omolayo}, \citenamefont {Paulic},
  \citenamefont {Ratti}, \citenamefont {Robin}, \citenamefont {Sun},
  \citenamefont {Swenson}, \citenamefont {Trovati}, \citenamefont {Venturini},
  \citenamefont {Waldron}, \citenamefont {Wallen},\ and\ \citenamefont
  {Wang}}]{alsu2025_10}%
  \BibitemOpen
  \bibfield  {author} {\bibinfo {author} {\bibfnamefont {C.}~\bibnamefont
  {Steier}}, \bibinfo {author} {\bibfnamefont {J.}~\bibnamefont {Bohon}},
  \bibinfo {author} {\bibfnamefont {S.}~\bibnamefont {Borra}}, \bibinfo
  {author} {\bibfnamefont {K.}~\bibnamefont {Chow}}, \bibinfo {author}
  {\bibfnamefont {C.}~\bibnamefont {Espino-Devine}}, \bibinfo {author}
  {\bibfnamefont {E.}~\bibnamefont {DiMasi}}, \bibinfo {author} {\bibfnamefont
  {G.}~\bibnamefont {Ganetis}}, \bibinfo {author} {\bibfnamefont
  {T.}~\bibnamefont {Hellert}}, \bibinfo {author} {\bibfnamefont
  {M.}~\bibnamefont {Johansson}}, \bibinfo {author} {\bibfnamefont
  {J.}~\bibnamefont {Joseph}}, \bibinfo {author} {\bibfnamefont {J.-Y.}\
  \bibnamefont {Jung}}, \bibinfo {author} {\bibfnamefont {R.}~\bibnamefont
  {Leftwich-Vann}}, \bibinfo {author} {\bibfnamefont {D.}~\bibnamefont
  {Leitner}}, \bibinfo {author} {\bibfnamefont {H.}~\bibnamefont {Lee}},
  \bibinfo {author} {\bibfnamefont {A.}~\bibnamefont {Lodge}}, \bibinfo
  {author} {\bibfnamefont {M.}~\bibnamefont {Lerche}}, \bibinfo {author}
  {\bibfnamefont {T.}~\bibnamefont {Luo}}, \bibinfo {author} {\bibfnamefont
  {R.}~\bibnamefont {Miller}}, \bibinfo {author} {\bibfnamefont
  {D.}~\bibnamefont {Nett}}, \bibinfo {author} {\bibfnamefont {B.}~\bibnamefont
  {Nicquevert}}, \bibinfo {author} {\bibfnamefont {O.}~\bibnamefont {Omolayo}},
  \bibinfo {author} {\bibfnamefont {D.}~\bibnamefont {Paulic}}, \bibinfo
  {author} {\bibfnamefont {A.}~\bibnamefont {Ratti}}, \bibinfo {author}
  {\bibfnamefont {D.}~\bibnamefont {Robin}}, \bibinfo {author} {\bibfnamefont
  {C.}~\bibnamefont {Sun}}, \bibinfo {author} {\bibfnamefont {C.}~\bibnamefont
  {Swenson}}, \bibinfo {author} {\bibfnamefont {S.}~\bibnamefont {Trovati}},
  \bibinfo {author} {\bibfnamefont {M.}~\bibnamefont {Venturini}}, \bibinfo
  {author} {\bibfnamefont {W.}~\bibnamefont {Waldron}}, \bibinfo {author}
  {\bibfnamefont {E.}~\bibnamefont {Wallen}},\ and\ \bibinfo {author}
  {\bibfnamefont {D.}~\bibnamefont {Wang}},\ }\href
  {https://doi.org/10.1088/1742-6596/3010/1/012046} {\bibfield  {journal}
  {\bibinfo  {journal} {J. Phys. Conf. Ser.}\ }\textbf {\bibinfo {volume}
  {3010}},\ \bibinfo {pages} {012046} (\bibinfo {year} {2025})}\BibitemShut
  {NoStop}%
\bibitem [{\citenamefont {Schroer}\ \emph {et~al.}(2018)\citenamefont
  {Schroer}, \citenamefont {Agapov}, \citenamefont {Brefeld}, \citenamefont
  {Brinkmann}, \citenamefont {Chae}, \citenamefont {Chao}, \citenamefont
  {Eriksson}, \citenamefont {Keil}, \citenamefont {Nuel~Gavald{\`a}},
  \citenamefont {R{\"o}hlsberger}, \citenamefont {Seeck}, \citenamefont
  {Sprung}, \citenamefont {Tischer}, \citenamefont {Wanzenberg},\ and\
  \citenamefont {Weckert}}]{petraiv_schroer_2018}%
  \BibitemOpen
  \bibfield  {author} {\bibinfo {author} {\bibfnamefont {C.~G.}\ \bibnamefont
  {Schroer}}, \bibinfo {author} {\bibfnamefont {I.}~\bibnamefont {Agapov}},
  \bibinfo {author} {\bibfnamefont {W.}~\bibnamefont {Brefeld}}, \bibinfo
  {author} {\bibfnamefont {R.}~\bibnamefont {Brinkmann}}, \bibinfo {author}
  {\bibfnamefont {Y.-C.}\ \bibnamefont {Chae}}, \bibinfo {author}
  {\bibfnamefont {H.-C.}\ \bibnamefont {Chao}}, \bibinfo {author}
  {\bibfnamefont {M.}~\bibnamefont {Eriksson}}, \bibinfo {author}
  {\bibfnamefont {J.}~\bibnamefont {Keil}}, \bibinfo {author} {\bibfnamefont
  {X.}~\bibnamefont {Nuel~Gavald{\`a}}}, \bibinfo {author} {\bibfnamefont
  {R.}~\bibnamefont {R{\"o}hlsberger}}, \bibinfo {author} {\bibfnamefont
  {O.~H.}\ \bibnamefont {Seeck}}, \bibinfo {author} {\bibfnamefont
  {M.}~\bibnamefont {Sprung}}, \bibinfo {author} {\bibfnamefont
  {M.}~\bibnamefont {Tischer}}, \bibinfo {author} {\bibfnamefont
  {R.}~\bibnamefont {Wanzenberg}},\ and\ \bibinfo {author} {\bibfnamefont
  {E.}~\bibnamefont {Weckert}},\ }\href
  {https://doi.org/10.1107/S1600577518008858} {\bibfield  {journal} {\bibinfo
  {journal} {J. Synchrotron Radiat.}\ }\textbf {\bibinfo {volume} {25}},\
  \bibinfo {pages} {1277} (\bibinfo {year} {2018})}\BibitemShut {NoStop}%
\bibitem [{\citenamefont {Borland}\ \emph {et~al.}(2014)\citenamefont
  {Borland}, \citenamefont {Decker}, \citenamefont {Emery}, \citenamefont
  {Sajaev}, \citenamefont {Sun},\ and\ \citenamefont
  {Xiao}}]{borland_lattice_challenges_2014}%
  \BibitemOpen
  \bibfield  {author} {\bibinfo {author} {\bibfnamefont {M.}~\bibnamefont
  {Borland}}, \bibinfo {author} {\bibfnamefont {G.}~\bibnamefont {Decker}},
  \bibinfo {author} {\bibfnamefont {L.}~\bibnamefont {Emery}}, \bibinfo
  {author} {\bibfnamefont {V.}~\bibnamefont {Sajaev}}, \bibinfo {author}
  {\bibfnamefont {Y.}~\bibnamefont {Sun}},\ and\ \bibinfo {author}
  {\bibfnamefont {A.}~\bibnamefont {Xiao}},\ }\href
  {https://doi.org/10.1107/S1600577514015203} {\bibfield  {journal} {\bibinfo
  {journal} {J. Synchrotron Radiat.}\ }\textbf {\bibinfo {volume} {21}},\
  \bibinfo {pages} {912} (\bibinfo {year} {2014})}\BibitemShut {NoStop}%
\bibitem [{\citenamefont {Sajaev}(2019)}]{sajaev_apsu_2019}%
  \BibitemOpen
  \bibfield  {author} {\bibinfo {author} {\bibfnamefont {V.}~\bibnamefont
  {Sajaev}},\ }\href {https://doi.org/10.1103/PhysRevAccelBeams.22.040102}
  {\bibfield  {journal} {\bibinfo  {journal} {Phys. Rev. Accel. Beams}\
  }\textbf {\bibinfo {volume} {22}},\ \bibinfo {pages} {040102} (\bibinfo
  {year} {2019})}\BibitemShut {NoStop}%
\bibitem [{\citenamefont {Hellert}\ \emph {et~al.}(2019)\citenamefont
  {Hellert}, \citenamefont {Amstutz}, \citenamefont {Steier},\ and\
  \citenamefont {Venturini}}]{hellert_sc_2019}%
  \BibitemOpen
  \bibfield  {author} {\bibinfo {author} {\bibfnamefont {T.}~\bibnamefont
  {Hellert}}, \bibinfo {author} {\bibfnamefont {P.}~\bibnamefont {Amstutz}},
  \bibinfo {author} {\bibfnamefont {C.}~\bibnamefont {Steier}},\ and\ \bibinfo
  {author} {\bibfnamefont {M.}~\bibnamefont {Venturini}},\ }\href
  {https://doi.org/10.1103/PhysRevAccelBeams.22.100702} {\bibfield  {journal}
  {\bibinfo  {journal} {Phys. Rev. Accel. Beams}\ }\textbf {\bibinfo {volume}
  {22}},\ \bibinfo {pages} {100702} (\bibinfo {year} {2019})}\BibitemShut
  {NoStop}%
\bibitem [{\citenamefont {Apollonio}\ \emph {et~al.}(2021)\citenamefont
  {Apollonio}, \citenamefont {Fielder}, \citenamefont {Ghasem},\ and\
  \citenamefont {Martin}}]{diamondii2021_33}%
  \BibitemOpen
  \bibfield  {author} {\bibinfo {author} {\bibfnamefont {M.}~\bibnamefont
  {Apollonio}}, \bibinfo {author} {\bibfnamefont {R.}~\bibnamefont {Fielder}},
  \bibinfo {author} {\bibfnamefont {H.}~\bibnamefont {Ghasem}},\ and\ \bibinfo
  {author} {\bibfnamefont {I.}~\bibnamefont {Martin}},\ }in\ \href
  {https://doi.org/10.18429/JACOW-IPAC2021-MOPAB063} {\emph {\bibinfo
  {booktitle} {Proc. IPAC'21}}}\ (\bibinfo  {publisher} {JACoW Publishing,
  Geneva, Switzerland},\ \bibinfo {year} {2021})\ pp.\ \bibinfo {pages}
  {265--268}\BibitemShut {NoStop}%
\bibitem [{\citenamefont {Amorim}\ \emph {et~al.}(2021)\citenamefont {Amorim},
  \citenamefont {Loulergue}, \citenamefont {Nadolski},\ and\ \citenamefont
  {Nagaoka}}]{soleilii2021_36}%
  \BibitemOpen
  \bibfield  {author} {\bibinfo {author} {\bibfnamefont {D.}~\bibnamefont
  {Amorim}}, \bibinfo {author} {\bibfnamefont {A.}~\bibnamefont {Loulergue}},
  \bibinfo {author} {\bibfnamefont {L.~S.}\ \bibnamefont {Nadolski}},\ and\
  \bibinfo {author} {\bibfnamefont {R.}~\bibnamefont {Nagaoka}},\ }in\ \href
  {https://doi.org/10.18429/JACOW-IPAC2021-MOPAB038} {\emph {\bibinfo
  {booktitle} {Proc. IPAC'21}}}\ (\bibinfo  {publisher} {JACoW Publishing,
  Geneva, Switzerland},\ \bibinfo {year} {2021})\ pp.\ \bibinfo {pages}
  {171--174}\BibitemShut {NoStop}%
\bibitem [{\citenamefont {Blanco-García}\ \emph {et~al.}(2022)\citenamefont
  {Blanco-García}, \citenamefont {Amorim}, \citenamefont {Deniaud},
  \citenamefont {Loulergue}, \citenamefont {Nadolski},\ and\ \citenamefont
  {Nagaoka}}]{soleilii2022_32}%
  \BibitemOpen
  \bibfield  {author} {\bibinfo {author} {\bibfnamefont {O.}~\bibnamefont
  {Blanco-García}}, \bibinfo {author} {\bibfnamefont {D.}~\bibnamefont
  {Amorim}}, \bibinfo {author} {\bibfnamefont {M.}~\bibnamefont {Deniaud}},
  \bibinfo {author} {\bibfnamefont {A.}~\bibnamefont {Loulergue}}, \bibinfo
  {author} {\bibfnamefont {L.}~\bibnamefont {Nadolski}},\ and\ \bibinfo
  {author} {\bibfnamefont {R.}~\bibnamefont {Nagaoka}},\ }in\ \href
  {https://doi.org/10.18429/JACoW-IPAC2022-MOPOTK004} {\emph {\bibinfo
  {booktitle} {Proc. IPAC'22}}}\ (\bibinfo  {publisher} {JACoW Publishing,
  Geneva, Switzerland},\ \bibinfo {year} {2022})\ pp.\ \bibinfo {pages}
  {433--436}\BibitemShut {NoStop}%
\bibitem [{\citenamefont {Hellert}\ \emph {et~al.}(2022)\citenamefont
  {Hellert}, \citenamefont {Agapov}, \citenamefont {Antipov}, \citenamefont
  {Bartolini}, \citenamefont {Brinkmann}, \citenamefont {Chae}, \citenamefont
  {Einfeld}, \citenamefont {Jebramcik},\ and\ \citenamefont
  {Keil}}]{petraiv2022_30}%
  \BibitemOpen
  \bibfield  {author} {\bibinfo {author} {\bibfnamefont {T.}~\bibnamefont
  {Hellert}}, \bibinfo {author} {\bibfnamefont {I.}~\bibnamefont {Agapov}},
  \bibinfo {author} {\bibfnamefont {S.}~\bibnamefont {Antipov}}, \bibinfo
  {author} {\bibfnamefont {R.}~\bibnamefont {Bartolini}}, \bibinfo {author}
  {\bibfnamefont {R.}~\bibnamefont {Brinkmann}}, \bibinfo {author}
  {\bibfnamefont {Y.-C.}\ \bibnamefont {Chae}}, \bibinfo {author}
  {\bibfnamefont {D.}~\bibnamefont {Einfeld}}, \bibinfo {author} {\bibfnamefont
  {M.}~\bibnamefont {Jebramcik}},\ and\ \bibinfo {author} {\bibfnamefont
  {J.}~\bibnamefont {Keil}},\ }in\ \href
  {https://doi.org/10.18429/JACoW-IPAC2022-TUPOMS018} {\emph {\bibinfo
  {booktitle} {Proc. IPAC'22}}}\ (\bibinfo  {publisher} {JACoW Publishing,
  Geneva, Switzerland},\ \bibinfo {year} {2022})\ pp.\ \bibinfo {pages}
  {1442--1444}\BibitemShut {NoStop}%
\bibitem [{\citenamefont {Habet}\ \emph {et~al.}(2025)\citenamefont {Habet},
  \citenamefont {Loulergue}, \citenamefont {Nadolski}, \citenamefont
  {Brunelle},\ and\ \citenamefont {Ducourtieux}}]{soleilii2025_9}%
  \BibitemOpen
  \bibfield  {author} {\bibinfo {author} {\bibfnamefont {S.}~\bibnamefont
  {Habet}}, \bibinfo {author} {\bibfnamefont {A.}~\bibnamefont {Loulergue}},
  \bibinfo {author} {\bibfnamefont {L.}~\bibnamefont {Nadolski}}, \bibinfo
  {author} {\bibfnamefont {P.}~\bibnamefont {Brunelle}},\ and\ \bibinfo
  {author} {\bibfnamefont {S.}~\bibnamefont {Ducourtieux}},\ }in\ \href
  {https://doi.org/10.18429/JACoW-IPAC2025-MOPS040} {\emph {\bibinfo
  {booktitle} {Proc. IPAC'25}}}\ (\bibinfo  {publisher} {JACoW Publishing,
  Geneva, Switzerland},\ \bibinfo {year} {2025})\ pp.\ \bibinfo {pages}
  {698--701}\BibitemShut {NoStop}%
\bibitem [{\citenamefont {Chen}\ \emph {et~al.}(2024)\citenamefont {Chen},
  \citenamefont {Wang}, \citenamefont {Wang}, \citenamefont {He}, \citenamefont
  {Wang}, \citenamefont {He}, \citenamefont {Hosaka},\ and\ \citenamefont
  {Xu}}]{half2024_15}%
  \BibitemOpen
  \bibfield  {author} {\bibinfo {author} {\bibfnamefont {K.}~\bibnamefont
  {Chen}}, \bibinfo {author} {\bibfnamefont {Z.}~\bibnamefont {Wang}}, \bibinfo
  {author} {\bibfnamefont {G.}~\bibnamefont {Wang}}, \bibinfo {author}
  {\bibfnamefont {T.}~\bibnamefont {He}}, \bibinfo {author} {\bibfnamefont
  {Z.}~\bibnamefont {Wang}}, \bibinfo {author} {\bibfnamefont {D.}~\bibnamefont
  {He}}, \bibinfo {author} {\bibfnamefont {M.}~\bibnamefont {Hosaka}},\ and\
  \bibinfo {author} {\bibfnamefont {W.}~\bibnamefont {Xu}},\ }in\ \href
  {https://doi.org/10.18429/JACoW-IPAC2024-WEPC38} {\emph {\bibinfo {booktitle}
  {Proc. IPAC'24}}}\ (\bibinfo  {publisher} {JACoW Publishing, Geneva,
  Switzerland},\ \bibinfo {year} {2024})\ pp.\ \bibinfo {pages}
  {2043--2045}\BibitemShut {NoStop}%
\bibitem [{\citenamefont {Benedetti}\ \emph {et~al.}(2025)\citenamefont
  {Benedetti}, \citenamefont {Perez}, \citenamefont {Carlà}, \citenamefont
  {Blanco-García},\ and\ \citenamefont {Martí}}]{albaii2025_8}%
  \BibitemOpen
  \bibfield  {author} {\bibinfo {author} {\bibfnamefont {G.}~\bibnamefont
  {Benedetti}}, \bibinfo {author} {\bibfnamefont {F.}~\bibnamefont {Perez}},
  \bibinfo {author} {\bibfnamefont {M.}~\bibnamefont {Carlà}}, \bibinfo
  {author} {\bibfnamefont {O.}~\bibnamefont {Blanco-García}},\ and\ \bibinfo
  {author} {\bibfnamefont {Z.}~\bibnamefont {Martí}},\ }in\ \href
  {https://doi.org/10.18429/JACoW-IPAC2025-WEPM105} {\emph {\bibinfo
  {booktitle} {Proc. IPAC'25}}}\ (\bibinfo  {publisher} {JACoW Publishing,
  Geneva, Switzerland},\ \bibinfo {year} {2025})\ pp.\ \bibinfo {pages}
  {2222--2224}\BibitemShut {NoStop}%
\bibitem [{\citenamefont {Wu}\ \emph {et~al.}(2024)\citenamefont {Wu},
  \citenamefont {Tan}, \citenamefont {Xuan}, \citenamefont {Tian},
  \citenamefont {Liu},\ and\ \citenamefont {Gong}}]{ssrfu_ipac2024}%
  \BibitemOpen
  \bibfield  {author} {\bibinfo {author} {\bibfnamefont {X.}~\bibnamefont
  {Wu}}, \bibinfo {author} {\bibfnamefont {L.}~\bibnamefont {Tan}}, \bibinfo
  {author} {\bibfnamefont {S.}~\bibnamefont {Xuan}}, \bibinfo {author}
  {\bibfnamefont {S.-Q.}\ \bibnamefont {Tian}}, \bibinfo {author}
  {\bibfnamefont {X.}~\bibnamefont {Liu}},\ and\ \bibinfo {author}
  {\bibfnamefont {Y.}~\bibnamefont {Gong}},\ }in\ \href
  {https://doi.org/10.18429/JACoW-IPAC2024-TUPG46} {\emph {\bibinfo {booktitle}
  {Proc. IPAC'24}}}\ (\bibinfo  {publisher} {JACoW Publishing, Geneva,
  Switzerland},\ \bibinfo {year} {2024})\ pp.\ \bibinfo {pages}
  {1342--1345}\BibitemShut {NoStop}%
\bibitem [{\citenamefont {Chao}\ and\ \citenamefont
  {Martin}(2024)}]{chao_diamondii_ipac2024}%
  \BibitemOpen
  \bibfield  {author} {\bibinfo {author} {\bibfnamefont {H.-C.}\ \bibnamefont
  {Chao}}\ and\ \bibinfo {author} {\bibfnamefont {I.}~\bibnamefont {Martin}},\
  }in\ \href {https://doi.org/10.18429/JACoW-IPAC2024-TUPG22} {\emph {\bibinfo
  {booktitle} {Proc. IPAC'24}}}\ (\bibinfo  {publisher} {JACoW Publishing,
  Geneva, Switzerland},\ \bibinfo {year} {2024})\ pp.\ \bibinfo {pages}
  {1262--1265}\BibitemShut {NoStop}%
\bibitem [{\citenamefont {Agapov}\ \emph {et~al.}(2024)\citenamefont {Agapov},
  \citenamefont {Antipov}, \citenamefont {Bartolini}, \citenamefont
  {Brinkmann}, \citenamefont {Chae}, \citenamefont {Cortes-Garcia},
  \citenamefont {Einfeld},\ and\ \citenamefont {Hellert}}]{petraiv2024_12}%
  \BibitemOpen
  \bibfield  {author} {\bibinfo {author} {\bibfnamefont {I.}~\bibnamefont
  {Agapov}}, \bibinfo {author} {\bibfnamefont {S.}~\bibnamefont {Antipov}},
  \bibinfo {author} {\bibfnamefont {R.}~\bibnamefont {Bartolini}}, \bibinfo
  {author} {\bibfnamefont {R.}~\bibnamefont {Brinkmann}}, \bibinfo {author}
  {\bibfnamefont {Y.}~\bibnamefont {Chae}}, \bibinfo {author} {\bibfnamefont
  {E.~C.}\ \bibnamefont {Cortes-Garcia}}, \bibinfo {author} {\bibfnamefont
  {D.}~\bibnamefont {Einfeld}},\ and\ \bibinfo {author} {\bibfnamefont
  {T.}~\bibnamefont {Hellert}},\ }\href@noop {} {\bibfield  {journal} {\bibinfo
   {journal} {arXiv preprint}\ } (\bibinfo {year} {2024})},\ \Eprint
  {https://arxiv.org/abs/2408.07995} {arXiv:2408.07995} \BibitemShut {NoStop}%
\bibitem [{\citenamefont {Martí}\ \emph {et~al.}(2023)\citenamefont {Martí},
  \citenamefont {Benedetti}, \citenamefont {Carlà},\ and\ \citenamefont
  {Iriso}}]{albaii2023_23}%
  \BibitemOpen
  \bibfield  {author} {\bibinfo {author} {\bibfnamefont {Z.}~\bibnamefont
  {Martí}}, \bibinfo {author} {\bibfnamefont {G.}~\bibnamefont {Benedetti}},
  \bibinfo {author} {\bibfnamefont {M.}~\bibnamefont {Carlà}},\ and\ \bibinfo
  {author} {\bibfnamefont {U.}~\bibnamefont {Iriso}},\ }in\ \href
  {https://doi.org/10.18429/JACoW-IPAC2023-WEPL003} {\emph {\bibinfo
  {booktitle} {Proc. IPAC'23}}}\ (\bibinfo  {publisher} {JACoW Publishing,
  Geneva, Switzerland},\ \bibinfo {year} {2023})\ pp.\ \bibinfo {pages}
  {3104--3107}\BibitemShut {NoStop}%
\bibitem [{\citenamefont {Ji}\ \emph {et~al.}(2023)\citenamefont {Ji},
  \citenamefont {Wang},\ and\ \citenamefont {Cui}}]{heps2023_29}%
  \BibitemOpen
  \bibfield  {author} {\bibinfo {author} {\bibfnamefont {D.}~\bibnamefont
  {Ji}}, \bibinfo {author} {\bibfnamefont {B.}~\bibnamefont {Wang}},\ and\
  \bibinfo {author} {\bibfnamefont {X.}~\bibnamefont {Cui}},\ }in\ \href
  {https://doi.org/10.18429/JACoW-IPAC2023-MOPA187} {\emph {\bibinfo
  {booktitle} {Proc. IPAC'23}}}\ (\bibinfo  {publisher} {JACoW Publishing,
  Geneva, Switzerland},\ \bibinfo {year} {2023})\ pp.\ \bibinfo {pages}
  {492--495}\BibitemShut {NoStop}%
\bibitem [{\citenamefont {Wang}\ \emph {et~al.}(2021)\citenamefont {Wang},
  \citenamefont {Duan}, \citenamefont {Ji}, \citenamefont {Jiao},\ and\
  \citenamefont {Zhao}}]{wang_heps_ipac2021}%
  \BibitemOpen
  \bibfield  {author} {\bibinfo {author} {\bibfnamefont {B.}~\bibnamefont
  {Wang}}, \bibinfo {author} {\bibfnamefont {Z.}~\bibnamefont {Duan}}, \bibinfo
  {author} {\bibfnamefont {D.}~\bibnamefont {Ji}}, \bibinfo {author}
  {\bibfnamefont {Y.}~\bibnamefont {Jiao}},\ and\ \bibinfo {author}
  {\bibfnamefont {Y.}~\bibnamefont {Zhao}},\ }in\ \href
  {https://doi.org/10.18429/JACoW-IPAC2021-TUPAB008} {\emph {\bibinfo
  {booktitle} {Proc. IPAC'21}}}\ (\bibinfo  {publisher} {JACoW Publishing,
  Geneva, Switzerland},\ \bibinfo {year} {2021})\ pp.\ \bibinfo {pages}
  {1349--1351}\BibitemShut {NoStop}%
\bibitem [{\citenamefont {Hellert}\ \emph {et~al.}(2023)\citenamefont
  {Hellert}, \citenamefont {Steier},\ and\ \citenamefont
  {Keil}}]{alsalsu2023_25}%
  \BibitemOpen
  \bibfield  {author} {\bibinfo {author} {\bibfnamefont {T.}~\bibnamefont
  {Hellert}}, \bibinfo {author} {\bibfnamefont {C.}~\bibnamefont {Steier}},\
  and\ \bibinfo {author} {\bibfnamefont {J.}~\bibnamefont {Keil}},\ }in\ \href
  {https://doi.org/10.18429/JACoW-IPAC2023-MOPM010} {\emph {\bibinfo
  {booktitle} {Proc. IPAC'23}}}\ (\bibinfo {year} {2023})\BibitemShut {NoStop}%
\bibitem [{\citenamefont {Huang}\ and\ \citenamefont
  {Safranek}(2015)}]{huang_online_2015}%
  \BibitemOpen
  \bibfield  {author} {\bibinfo {author} {\bibfnamefont {X.}~\bibnamefont
  {Huang}}\ and\ \bibinfo {author} {\bibfnamefont {J.}~\bibnamefont
  {Safranek}},\ }\href {https://doi.org/10.1103/PhysRevSTAB.18.084001}
  {\bibfield  {journal} {\bibinfo  {journal} {Phys. Rev. ST Accel. Beams}\
  }\textbf {\bibinfo {volume} {18}},\ \bibinfo {pages} {084001} (\bibinfo
  {year} {2015})},\ \Eprint {https://arxiv.org/abs/1502.07799}
  {arXiv:1502.07799} \BibitemShut {NoStop}%
\bibitem [{\citenamefont {Duris}\ \emph {et~al.}(2020)\citenamefont {Duris},
  \citenamefont {Kennedy}, \citenamefont {Hanuka}, \citenamefont {Shtalenkova},
  \citenamefont {Edelen}, \citenamefont {Baxevanis}, \citenamefont {Egger},
  \citenamefont {Cope}, \citenamefont {McIntire}, \citenamefont {Ermon},\ and\
  \citenamefont {Ratner}}]{duris_bo_fel_2020}%
  \BibitemOpen
  \bibfield  {author} {\bibinfo {author} {\bibfnamefont {J.}~\bibnamefont
  {Duris}}, \bibinfo {author} {\bibfnamefont {D.}~\bibnamefont {Kennedy}},
  \bibinfo {author} {\bibfnamefont {A.}~\bibnamefont {Hanuka}}, \bibinfo
  {author} {\bibfnamefont {J.}~\bibnamefont {Shtalenkova}}, \bibinfo {author}
  {\bibfnamefont {A.}~\bibnamefont {Edelen}}, \bibinfo {author} {\bibfnamefont
  {P.}~\bibnamefont {Baxevanis}}, \bibinfo {author} {\bibfnamefont
  {A.}~\bibnamefont {Egger}}, \bibinfo {author} {\bibfnamefont
  {T.}~\bibnamefont {Cope}}, \bibinfo {author} {\bibfnamefont {M.}~\bibnamefont
  {McIntire}}, \bibinfo {author} {\bibfnamefont {S.}~\bibnamefont {Ermon}},\
  and\ \bibinfo {author} {\bibfnamefont {D.}~\bibnamefont {Ratner}},\ }\href
  {https://doi.org/10.1103/PhysRevLett.124.124801} {\bibfield  {journal}
  {\bibinfo  {journal} {Phys. Rev. Lett.}\ }\textbf {\bibinfo {volume} {124}},\
  \bibinfo {pages} {124801} (\bibinfo {year} {2020})},\ \Eprint
  {https://arxiv.org/abs/1909.05963} {arXiv:1909.05963} \BibitemShut {NoStop}%
\bibitem [{\citenamefont {Roussel}\ \emph {et~al.}(2021)\citenamefont
  {Roussel}, \citenamefont {Hanuka},\ and\ \citenamefont
  {Edelen}}]{roussel_mobo_2021}%
  \BibitemOpen
  \bibfield  {author} {\bibinfo {author} {\bibfnamefont {R.}~\bibnamefont
  {Roussel}}, \bibinfo {author} {\bibfnamefont {A.}~\bibnamefont {Hanuka}},\
  and\ \bibinfo {author} {\bibfnamefont {A.}~\bibnamefont {Edelen}},\ }\href
  {https://doi.org/10.1103/PhysRevAccelBeams.24.062801} {\bibfield  {journal}
  {\bibinfo  {journal} {Phys. Rev. Accel. Beams}\ }\textbf {\bibinfo {volume}
  {24}},\ \bibinfo {pages} {062801} (\bibinfo {year} {2021})},\ \Eprint
  {https://arxiv.org/abs/2010.09824} {arXiv:2010.09824} \BibitemShut {NoStop}%
\bibitem [{\citenamefont {Xu}\ \emph {et~al.}(2023)\citenamefont {Xu},
  \citenamefont {Boltz}, \citenamefont {Mochihashi}, \citenamefont {Garcia},
  \citenamefont {Schuh},\ and\ \citenamefont
  {M{\"u}ller}}]{xu_bo_injection_2023}%
  \BibitemOpen
  \bibfield  {author} {\bibinfo {author} {\bibfnamefont {C.}~\bibnamefont
  {Xu}}, \bibinfo {author} {\bibfnamefont {T.}~\bibnamefont {Boltz}}, \bibinfo
  {author} {\bibfnamefont {A.}~\bibnamefont {Mochihashi}}, \bibinfo {author}
  {\bibfnamefont {A.~S.}\ \bibnamefont {Garcia}}, \bibinfo {author}
  {\bibfnamefont {M.}~\bibnamefont {Schuh}},\ and\ \bibinfo {author}
  {\bibfnamefont {A.-S.}\ \bibnamefont {M{\"u}ller}},\ }\href
  {https://doi.org/10.1103/PhysRevAccelBeams.26.034601} {\bibfield  {journal}
  {\bibinfo  {journal} {Phys. Rev. Accel. Beams}\ }\textbf {\bibinfo {volume}
  {26}},\ \bibinfo {pages} {034601} (\bibinfo {year} {2023})},\ \Eprint
  {https://arxiv.org/abs/2211.09504} {arXiv:2211.09504} \BibitemShut {NoStop}%
\bibitem [{\citenamefont {Kaiser}\ \emph
  {et~al.}(2024{\natexlab{a}})\citenamefont {Kaiser}, \citenamefont {Xu},
  \citenamefont {Eichler},\ and\ \citenamefont {Garcia}}]{kaiser_cheetah_2024}%
  \BibitemOpen
  \bibfield  {author} {\bibinfo {author} {\bibfnamefont {J.}~\bibnamefont
  {Kaiser}}, \bibinfo {author} {\bibfnamefont {C.}~\bibnamefont {Xu}}, \bibinfo
  {author} {\bibfnamefont {A.}~\bibnamefont {Eichler}},\ and\ \bibinfo {author}
  {\bibfnamefont {A.~S.}\ \bibnamefont {Garcia}},\ }\href
  {https://doi.org/10.1103/PhysRevAccelBeams.27.054601} {\bibfield  {journal}
  {\bibinfo  {journal} {Phys. Rev. Accel. Beams}\ }\textbf {\bibinfo {volume}
  {27}},\ \bibinfo {pages} {054601} (\bibinfo {year} {2024}{\natexlab{a}})},\
  \Eprint {https://arxiv.org/abs/2401.05815} {arXiv:2401.05815} \BibitemShut
  {NoStop}%
\bibitem [{\citenamefont {Kaiser}\ \emph
  {et~al.}(2024{\natexlab{b}})\citenamefont {Kaiser}, \citenamefont {Xu},
  \citenamefont {Eichler}, \citenamefont {Garcia}, \citenamefont {Stein},
  \citenamefont {Br{\"u}ndermann}, \citenamefont {Kuropka}, \citenamefont
  {Dinter}, \citenamefont {Mayet}, \citenamefont {Vinatier}, \citenamefont
  {Burkart},\ and\ \citenamefont {Schlarb}}]{kaiser_rl_vs_bo_2024}%
  \BibitemOpen
  \bibfield  {author} {\bibinfo {author} {\bibfnamefont {J.}~\bibnamefont
  {Kaiser}}, \bibinfo {author} {\bibfnamefont {C.}~\bibnamefont {Xu}}, \bibinfo
  {author} {\bibfnamefont {A.}~\bibnamefont {Eichler}}, \bibinfo {author}
  {\bibfnamefont {A.~S.}\ \bibnamefont {Garcia}}, \bibinfo {author}
  {\bibfnamefont {O.}~\bibnamefont {Stein}}, \bibinfo {author} {\bibfnamefont
  {E.}~\bibnamefont {Br{\"u}ndermann}}, \bibinfo {author} {\bibfnamefont
  {W.}~\bibnamefont {Kuropka}}, \bibinfo {author} {\bibfnamefont
  {H.}~\bibnamefont {Dinter}}, \bibinfo {author} {\bibfnamefont
  {F.}~\bibnamefont {Mayet}}, \bibinfo {author} {\bibfnamefont
  {T.}~\bibnamefont {Vinatier}}, \bibinfo {author} {\bibfnamefont
  {F.}~\bibnamefont {Burkart}},\ and\ \bibinfo {author} {\bibfnamefont
  {H.}~\bibnamefont {Schlarb}},\ }\href
  {https://doi.org/10.1038/s41598-024-66263-y} {\bibfield  {journal} {\bibinfo
  {journal} {Sci. Rep.}\ }\textbf {\bibinfo {volume} {14}},\ \bibinfo {pages}
  {15733} (\bibinfo {year} {2024}{\natexlab{b}})},\ \Eprint
  {https://arxiv.org/abs/2306.03739} {arXiv:2306.03739} \BibitemShut {NoStop}%
\bibitem [{\citenamefont {Roussel}\ \emph {et~al.}(2024)\citenamefont
  {Roussel}, \citenamefont {Edelen}, \citenamefont {Boltz}, \citenamefont
  {Kennedy}, \citenamefont {Zhang}, \citenamefont {Ji}, \citenamefont {Huang},
  \citenamefont {Ratner}, \citenamefont {Garcia}, \citenamefont {Xu},
  \citenamefont {Kaiser}, \citenamefont {Pousa}, \citenamefont {Eichler},
  \citenamefont {Lübsen}, \citenamefont {Isenberg}, \citenamefont {Gao},
  \citenamefont {Kuklev}, \citenamefont {Martinez}, \citenamefont {Mustapha},
  \citenamefont {Kain}, \citenamefont {Mayes}, \citenamefont {Lin},
  \citenamefont {Liuzzo}, \citenamefont {John}, \citenamefont {Streeter},
  \citenamefont {Lehe},\ and\ \citenamefont
  {Neiswanger}}]{roussel_bo_review_2024}%
  \BibitemOpen
  \bibfield  {author} {\bibinfo {author} {\bibfnamefont {R.}~\bibnamefont
  {Roussel}}, \bibinfo {author} {\bibfnamefont {A.~L.}\ \bibnamefont {Edelen}},
  \bibinfo {author} {\bibfnamefont {T.}~\bibnamefont {Boltz}}, \bibinfo
  {author} {\bibfnamefont {D.}~\bibnamefont {Kennedy}}, \bibinfo {author}
  {\bibfnamefont {Z.}~\bibnamefont {Zhang}}, \bibinfo {author} {\bibfnamefont
  {F.}~\bibnamefont {Ji}}, \bibinfo {author} {\bibfnamefont {X.}~\bibnamefont
  {Huang}}, \bibinfo {author} {\bibfnamefont {D.}~\bibnamefont {Ratner}},
  \bibinfo {author} {\bibfnamefont {A.~S.}\ \bibnamefont {Garcia}}, \bibinfo
  {author} {\bibfnamefont {C.}~\bibnamefont {Xu}}, \bibinfo {author}
  {\bibfnamefont {J.}~\bibnamefont {Kaiser}}, \bibinfo {author} {\bibfnamefont
  {A.~F.}\ \bibnamefont {Pousa}}, \bibinfo {author} {\bibfnamefont
  {A.}~\bibnamefont {Eichler}}, \bibinfo {author} {\bibfnamefont {J.~O.}\
  \bibnamefont {Lübsen}}, \bibinfo {author} {\bibfnamefont {N.~M.}\
  \bibnamefont {Isenberg}}, \bibinfo {author} {\bibfnamefont {Y.}~\bibnamefont
  {Gao}}, \bibinfo {author} {\bibfnamefont {N.}~\bibnamefont {Kuklev}},
  \bibinfo {author} {\bibfnamefont {J.}~\bibnamefont {Martinez}}, \bibinfo
  {author} {\bibfnamefont {B.}~\bibnamefont {Mustapha}}, \bibinfo {author}
  {\bibfnamefont {V.}~\bibnamefont {Kain}}, \bibinfo {author} {\bibfnamefont
  {C.}~\bibnamefont {Mayes}}, \bibinfo {author} {\bibfnamefont
  {W.}~\bibnamefont {Lin}}, \bibinfo {author} {\bibfnamefont {S.~M.}\
  \bibnamefont {Liuzzo}}, \bibinfo {author} {\bibfnamefont {J.~S.}\
  \bibnamefont {John}}, \bibinfo {author} {\bibfnamefont {M.~J.~V.}\
  \bibnamefont {Streeter}}, \bibinfo {author} {\bibfnamefont {R.}~\bibnamefont
  {Lehe}},\ and\ \bibinfo {author} {\bibfnamefont {W.}~\bibnamefont
  {Neiswanger}},\ }\href {https://doi.org/10.1103/PhysRevAccelBeams.27.084801}
  {\bibfield  {journal} {\bibinfo  {journal} {Phys. Rev. Accel. Beams}\
  }\textbf {\bibinfo {volume} {27}},\ \bibinfo {pages} {084801} (\bibinfo
  {year} {2024})},\ \Eprint {https://arxiv.org/abs/2312.05667}
  {arXiv:2312.05667} \BibitemShut {NoStop}%
\bibitem [{\citenamefont {Kaiser}\ \emph {et~al.}(2025)\citenamefont {Kaiser},
  \citenamefont {Lauscher},\ and\ \citenamefont
  {Eichler}}]{kaiser_llm_tuning_2025}%
  \BibitemOpen
  \bibfield  {author} {\bibinfo {author} {\bibfnamefont {J.}~\bibnamefont
  {Kaiser}}, \bibinfo {author} {\bibfnamefont {A.}~\bibnamefont {Lauscher}},\
  and\ \bibinfo {author} {\bibfnamefont {A.}~\bibnamefont {Eichler}},\ }\href
  {https://doi.org/10.1126/sciadv.adr4173} {\bibfield  {journal} {\bibinfo
  {journal} {Sci. Adv.}\ }\textbf {\bibinfo {volume} {11}},\ \bibinfo {pages}
  {eadr4173} (\bibinfo {year} {2025})},\ \Eprint
  {https://arxiv.org/abs/2405.08888} {arXiv:2405.08888} \BibitemShut {NoStop}%
\bibitem [{\citenamefont {Mayet}(2024)}]{mayet_gaia_2024}%
  \BibitemOpen
  \bibfield  {author} {\bibinfo {author} {\bibfnamefont {F.}~\bibnamefont
  {Mayet}},\ }\href@noop {} {\bibinfo {title} {{GAIA: A General AI Assistant
  for Intelligent Accelerator Operations}}} (\bibinfo {year} {2024}),\ \Eprint
  {https://arxiv.org/abs/2405.01359} {arXiv:2405.01359 [cs.CL]} \BibitemShut
  {NoStop}%
\bibitem [{\citenamefont {Sulc}\ \emph {et~al.}(2024)\citenamefont {Sulc},
  \citenamefont {Hellert}, \citenamefont {Kammering}, \citenamefont
  {Hoschouer},\ and\ \citenamefont {John}}]{sulc_agentic_2024}%
  \BibitemOpen
  \bibfield  {author} {\bibinfo {author} {\bibfnamefont {A.}~\bibnamefont
  {Sulc}}, \bibinfo {author} {\bibfnamefont {T.}~\bibnamefont {Hellert}},
  \bibinfo {author} {\bibfnamefont {R.}~\bibnamefont {Kammering}}, \bibinfo
  {author} {\bibfnamefont {H.}~\bibnamefont {Hoschouer}},\ and\ \bibinfo
  {author} {\bibfnamefont {J.~S.}\ \bibnamefont {John}},\ }in\ \href@noop {}
  {\emph {\bibinfo {booktitle} {Machine Learning and the Physical Sciences
  Workshop, NeurIPS 2024}}}\ (\bibinfo {year} {2024})\ \Eprint
  {https://arxiv.org/abs/2409.06336} {arXiv:2409.06336 [physics.acc-ph]}
  \BibitemShut {NoStop}%
\bibitem [{\citenamefont {Hellert}\ \emph {et~al.}(2026)\citenamefont
  {Hellert}, \citenamefont {Bertwistle}, \citenamefont {Leemann}, \citenamefont
  {Sulc},\ and\ \citenamefont {Venturini}}]{hellert_agentic_prr}%
  \BibitemOpen
  \bibfield  {author} {\bibinfo {author} {\bibfnamefont {T.}~\bibnamefont
  {Hellert}}, \bibinfo {author} {\bibfnamefont {D.}~\bibnamefont {Bertwistle}},
  \bibinfo {author} {\bibfnamefont {S.~C.}\ \bibnamefont {Leemann}}, \bibinfo
  {author} {\bibfnamefont {A.}~\bibnamefont {Sulc}},\ and\ \bibinfo {author}
  {\bibfnamefont {M.}~\bibnamefont {Venturini}},\ }\href
  {https://doi.org/10.1103/jtqy-9jz1} {\bibfield  {journal} {\bibinfo
  {journal} {Phys. Rev. Research}\ }\textbf {\bibinfo {volume} {8}},\ \bibinfo
  {pages} {L012017} (\bibinfo {year} {2026})},\ \Eprint
  {https://arxiv.org/abs/2509.17255} {arXiv:2509.17255} \BibitemShut {NoStop}%
\bibitem [{\citenamefont {Hellert}\ \emph {et~al.}(2025)\citenamefont
  {Hellert}, \citenamefont {Montenegro},\ and\ \citenamefont
  {Sulc}}]{osprey_2025}%
  \BibitemOpen
  \bibfield  {author} {\bibinfo {author} {\bibfnamefont {T.}~\bibnamefont
  {Hellert}}, \bibinfo {author} {\bibfnamefont {J.}~\bibnamefont
  {Montenegro}},\ and\ \bibinfo {author} {\bibfnamefont {A.}~\bibnamefont
  {Sulc}},\ }\href@noop {} {\bibinfo {title} {{Osprey: Production-Ready Agentic
  AI for Safety-Critical Control Systems}}} (\bibinfo {year} {2025}),\ \Eprint
  {https://arxiv.org/abs/2508.15066} {arXiv:2508.15066 [cs.MA]} \BibitemShut
  {NoStop}%
\bibitem [{\citenamefont {Karpathy}(2026)}]{karpathy_autoresearch_2026}%
  \BibitemOpen
  \bibfield  {author} {\bibinfo {author} {\bibfnamefont {A.}~\bibnamefont
  {Karpathy}},\ }\href {https://github.com/karpathy/autoresearch} {\bibinfo
  {title} {\texttt{autoresearch}: {AI} agents running research on single-{GPU}
  nanochat training automatically}},\ \bibinfo {howpublished}
  {\url{https://github.com/karpathy/autoresearch}} (\bibinfo {year} {2026}),\
  \bibinfo {note} {open-source greedy experiment loop for automated {ML}
  research}\BibitemShut {NoStop}%
\bibitem [{\citenamefont {Lu}\ \emph {et~al.}(2024)\citenamefont {Lu},
  \citenamefont {Lu}, \citenamefont {Lange}, \citenamefont {Foerster},
  \citenamefont {Clune},\ and\ \citenamefont {Ha}}]{lu_ai_scientist_2024}%
  \BibitemOpen
  \bibfield  {author} {\bibinfo {author} {\bibfnamefont {C.}~\bibnamefont
  {Lu}}, \bibinfo {author} {\bibfnamefont {C.}~\bibnamefont {Lu}}, \bibinfo
  {author} {\bibfnamefont {R.~T.}\ \bibnamefont {Lange}}, \bibinfo {author}
  {\bibfnamefont {J.}~\bibnamefont {Foerster}}, \bibinfo {author}
  {\bibfnamefont {J.}~\bibnamefont {Clune}},\ and\ \bibinfo {author}
  {\bibfnamefont {D.}~\bibnamefont {Ha}},\ }\href@noop {} {\bibinfo {title}
  {{The AI Scientist: Towards Fully Automated Open-Ended Scientific
  Discovery}}} (\bibinfo {year} {2024}),\ \Eprint
  {https://arxiv.org/abs/2408.06292} {arXiv:2408.06292 [cs.AI]} \BibitemShut
  {NoStop}%
\bibitem [{\citenamefont {Boiko}\ \emph {et~al.}(2023)\citenamefont {Boiko},
  \citenamefont {MacKnight}, \citenamefont {Kline},\ and\ \citenamefont
  {Gomes}}]{boiko_coscientist_2023}%
  \BibitemOpen
  \bibfield  {author} {\bibinfo {author} {\bibfnamefont {D.~A.}\ \bibnamefont
  {Boiko}}, \bibinfo {author} {\bibfnamefont {R.}~\bibnamefont {MacKnight}},
  \bibinfo {author} {\bibfnamefont {B.}~\bibnamefont {Kline}},\ and\ \bibinfo
  {author} {\bibfnamefont {G.}~\bibnamefont {Gomes}},\ }\href
  {https://doi.org/10.1038/s41586-023-06792-0} {\bibfield  {journal} {\bibinfo
  {journal} {Nature}\ }\textbf {\bibinfo {volume} {624}},\ \bibinfo {pages}
  {570} (\bibinfo {year} {2023})}\BibitemShut {NoStop}%
\bibitem [{\citenamefont {Bran}\ \emph {et~al.}(2024)\citenamefont {Bran},
  \citenamefont {Cox}, \citenamefont {Schilter}, \citenamefont {Baldassari},
  \citenamefont {White},\ and\ \citenamefont {Schwaller}}]{bran_chemcrow_2024}%
  \BibitemOpen
  \bibfield  {author} {\bibinfo {author} {\bibfnamefont {A.~M.}\ \bibnamefont
  {Bran}}, \bibinfo {author} {\bibfnamefont {S.}~\bibnamefont {Cox}}, \bibinfo
  {author} {\bibfnamefont {O.}~\bibnamefont {Schilter}}, \bibinfo {author}
  {\bibfnamefont {C.}~\bibnamefont {Baldassari}}, \bibinfo {author}
  {\bibfnamefont {A.~D.}\ \bibnamefont {White}},\ and\ \bibinfo {author}
  {\bibfnamefont {P.}~\bibnamefont {Schwaller}},\ }\href
  {https://doi.org/10.1038/s42256-024-00832-8} {\bibfield  {journal} {\bibinfo
  {journal} {Nat. Mach. Intell.}\ }\textbf {\bibinfo {volume} {6}},\ \bibinfo
  {pages} {525} (\bibinfo {year} {2024})},\ \Eprint
  {https://arxiv.org/abs/2304.05376} {arXiv:2304.05376} \BibitemShut {NoStop}%
\bibitem [{\citenamefont {Romera-Paredes}\ \emph {et~al.}(2024)\citenamefont
  {Romera-Paredes}, \citenamefont {Barekatain}, \citenamefont {Novikov},
  \citenamefont {Balog}, \citenamefont {Kumar}, \citenamefont {Dupont},
  \citenamefont {Ruiz}, \citenamefont {Ellenberg}, \citenamefont {Wang},
  \citenamefont {Fawzi}, \citenamefont {Kohli},\ and\ \citenamefont
  {Fawzi}}]{funsearch_2023}%
  \BibitemOpen
  \bibfield  {author} {\bibinfo {author} {\bibfnamefont {B.}~\bibnamefont
  {Romera-Paredes}}, \bibinfo {author} {\bibfnamefont {M.}~\bibnamefont
  {Barekatain}}, \bibinfo {author} {\bibfnamefont {A.}~\bibnamefont {Novikov}},
  \bibinfo {author} {\bibfnamefont {M.}~\bibnamefont {Balog}}, \bibinfo
  {author} {\bibfnamefont {M.~P.}\ \bibnamefont {Kumar}}, \bibinfo {author}
  {\bibfnamefont {E.}~\bibnamefont {Dupont}}, \bibinfo {author} {\bibfnamefont
  {F.~J.~R.}\ \bibnamefont {Ruiz}}, \bibinfo {author} {\bibfnamefont {J.~S.}\
  \bibnamefont {Ellenberg}}, \bibinfo {author} {\bibfnamefont {P.}~\bibnamefont
  {Wang}}, \bibinfo {author} {\bibfnamefont {O.}~\bibnamefont {Fawzi}},
  \bibinfo {author} {\bibfnamefont {P.}~\bibnamefont {Kohli}},\ and\ \bibinfo
  {author} {\bibfnamefont {A.}~\bibnamefont {Fawzi}},\ }\href
  {https://doi.org/10.1038/s41586-023-06924-6} {\bibfield  {journal} {\bibinfo
  {journal} {Nature}\ }\textbf {\bibinfo {volume} {625}},\ \bibinfo {pages}
  {468} (\bibinfo {year} {2024})}\BibitemShut {NoStop}%
\bibitem [{\citenamefont {Ma}\ \emph {et~al.}(2024)\citenamefont {Ma},
  \citenamefont {Liang}, \citenamefont {Wang}, \citenamefont {Huang},
  \citenamefont {Bastani}, \citenamefont {Jayaraman}, \citenamefont {Zhu},
  \citenamefont {Fan},\ and\ \citenamefont {Anandkumar}}]{ma_eureka_2024}%
  \BibitemOpen
  \bibfield  {author} {\bibinfo {author} {\bibfnamefont {Y.~J.}\ \bibnamefont
  {Ma}}, \bibinfo {author} {\bibfnamefont {W.}~\bibnamefont {Liang}}, \bibinfo
  {author} {\bibfnamefont {G.}~\bibnamefont {Wang}}, \bibinfo {author}
  {\bibfnamefont {D.-A.}\ \bibnamefont {Huang}}, \bibinfo {author}
  {\bibfnamefont {O.}~\bibnamefont {Bastani}}, \bibinfo {author} {\bibfnamefont
  {D.}~\bibnamefont {Jayaraman}}, \bibinfo {author} {\bibfnamefont
  {Y.}~\bibnamefont {Zhu}}, \bibinfo {author} {\bibfnamefont {L.}~\bibnamefont
  {Fan}},\ and\ \bibinfo {author} {\bibfnamefont {A.}~\bibnamefont
  {Anandkumar}},\ }in\ \href@noop {} {\emph {\bibinfo {booktitle} {Int. Conf.
  on Learning Representations (ICLR)}}}\ (\bibinfo {year} {2024})\ \Eprint
  {https://arxiv.org/abs/2310.12931} {arXiv:2310.12931 [cs.LG]} \BibitemShut
  {NoStop}%
\bibitem [{\citenamefont {MacLeod}\ \emph {et~al.}(2020)\citenamefont
  {MacLeod}, \citenamefont {Parlane}, \citenamefont {Morrissey}, \citenamefont
  {H{\"a}se}, \citenamefont {Roch}, \citenamefont {Dettelbach}, \citenamefont
  {Moreira}, \citenamefont {Yunker}, \citenamefont {Rooney}, \citenamefont
  {Deeth}, \citenamefont {Lai}, \citenamefont {Ng}, \citenamefont {Situ},
  \citenamefont {Zhang}, \citenamefont {Elliott}, \citenamefont {Haley},
  \citenamefont {Dvorak}, \citenamefont {Aspuru-Guzik}, \citenamefont {Hein},\
  and\ \citenamefont {Berlinguette}}]{macleod_sdl_2020}%
  \BibitemOpen
  \bibfield  {author} {\bibinfo {author} {\bibfnamefont {B.~P.}\ \bibnamefont
  {MacLeod}}, \bibinfo {author} {\bibfnamefont {F.~G.~L.}\ \bibnamefont
  {Parlane}}, \bibinfo {author} {\bibfnamefont {T.~D.}\ \bibnamefont
  {Morrissey}}, \bibinfo {author} {\bibfnamefont {F.}~\bibnamefont {H{\"a}se}},
  \bibinfo {author} {\bibfnamefont {L.~M.}\ \bibnamefont {Roch}}, \bibinfo
  {author} {\bibfnamefont {K.~E.}\ \bibnamefont {Dettelbach}}, \bibinfo
  {author} {\bibfnamefont {R.}~\bibnamefont {Moreira}}, \bibinfo {author}
  {\bibfnamefont {L.~P.~E.}\ \bibnamefont {Yunker}}, \bibinfo {author}
  {\bibfnamefont {M.~B.}\ \bibnamefont {Rooney}}, \bibinfo {author}
  {\bibfnamefont {J.~R.}\ \bibnamefont {Deeth}}, \bibinfo {author}
  {\bibfnamefont {V.}~\bibnamefont {Lai}}, \bibinfo {author} {\bibfnamefont
  {G.~J.}\ \bibnamefont {Ng}}, \bibinfo {author} {\bibfnamefont
  {H.}~\bibnamefont {Situ}}, \bibinfo {author} {\bibfnamefont {R.~H.}\
  \bibnamefont {Zhang}}, \bibinfo {author} {\bibfnamefont {M.~S.}\ \bibnamefont
  {Elliott}}, \bibinfo {author} {\bibfnamefont {T.~H.}\ \bibnamefont {Haley}},
  \bibinfo {author} {\bibfnamefont {D.~J.}\ \bibnamefont {Dvorak}}, \bibinfo
  {author} {\bibfnamefont {A.}~\bibnamefont {Aspuru-Guzik}}, \bibinfo {author}
  {\bibfnamefont {J.~E.}\ \bibnamefont {Hein}},\ and\ \bibinfo {author}
  {\bibfnamefont {C.~P.}\ \bibnamefont {Berlinguette}},\ }\href
  {https://doi.org/10.1126/sciadv.aaz8867} {\bibfield  {journal} {\bibinfo
  {journal} {Sci. Adv.}\ }\textbf {\bibinfo {volume} {6}},\ \bibinfo {pages}
  {eaaz8867} (\bibinfo {year} {2020})}\BibitemShut {NoStop}%
\bibitem [{\citenamefont {Szymanski}\ \emph {et~al.}(2023)\citenamefont
  {Szymanski}, \citenamefont {Rendy}, \citenamefont {Fei}, \citenamefont
  {Kumar}, \citenamefont {He}, \citenamefont {Milsted} \emph
  {et~al.}}]{szymanski_alab_2023}%
  \BibitemOpen
  \bibfield  {author} {\bibinfo {author} {\bibfnamefont {N.~J.}\ \bibnamefont
  {Szymanski}}, \bibinfo {author} {\bibfnamefont {B.}~\bibnamefont {Rendy}},
  \bibinfo {author} {\bibfnamefont {Y.}~\bibnamefont {Fei}}, \bibinfo {author}
  {\bibfnamefont {R.~E.}\ \bibnamefont {Kumar}}, \bibinfo {author}
  {\bibfnamefont {T.}~\bibnamefont {He}}, \bibinfo {author} {\bibfnamefont
  {D.}~\bibnamefont {Milsted}}, \emph {et~al.},\ }\href
  {https://doi.org/10.1038/s41586-023-06734-w} {\bibfield  {journal} {\bibinfo
  {journal} {Nature}\ }\textbf {\bibinfo {volume} {624}},\ \bibinfo {pages}
  {86} (\bibinfo {year} {2023})}\BibitemShut {NoStop}%
\bibitem [{\citenamefont {Abolhasani}\ and\ \citenamefont
  {Kumacheva}(2023)}]{abolhasani_sdl_2023}%
  \BibitemOpen
  \bibfield  {author} {\bibinfo {author} {\bibfnamefont {M.}~\bibnamefont
  {Abolhasani}}\ and\ \bibinfo {author} {\bibfnamefont {E.}~\bibnamefont
  {Kumacheva}},\ }\href {https://doi.org/10.1038/s44160-022-00231-0} {\bibfield
   {journal} {\bibinfo  {journal} {Nat. Synth.}\ }\textbf {\bibinfo {volume}
  {2}},\ \bibinfo {pages} {483} (\bibinfo {year} {2023})}\BibitemShut {NoStop}%
\bibitem [{\citenamefont {Stach}\ \emph {et~al.}(2021)\citenamefont {Stach},
  \citenamefont {DeCost}, \citenamefont {Kusne}, \citenamefont
  {Hattrick-Simpers}, \citenamefont {Brown}, \citenamefont {Reyes} \emph
  {et~al.}}]{stach_ae_2021}%
  \BibitemOpen
  \bibfield  {author} {\bibinfo {author} {\bibfnamefont {E.}~\bibnamefont
  {Stach}}, \bibinfo {author} {\bibfnamefont {B.}~\bibnamefont {DeCost}},
  \bibinfo {author} {\bibfnamefont {A.~G.}\ \bibnamefont {Kusne}}, \bibinfo
  {author} {\bibfnamefont {J.}~\bibnamefont {Hattrick-Simpers}}, \bibinfo
  {author} {\bibfnamefont {K.~A.}\ \bibnamefont {Brown}}, \bibinfo {author}
  {\bibfnamefont {K.~G.}\ \bibnamefont {Reyes}}, \emph {et~al.},\ }\href
  {https://doi.org/10.1016/j.matt.2021.06.036} {\bibfield  {journal} {\bibinfo
  {journal} {Matter}\ }\textbf {\bibinfo {volume} {4}},\ \bibinfo {pages}
  {2702} (\bibinfo {year} {2021})}\BibitemShut {NoStop}%
\bibitem [{\citenamefont {Steier}\ \emph {et~al.}(2019)\citenamefont {Steier},
  \citenamefont {Amstutz}, \citenamefont {Baptiste}, \citenamefont {Bong},
  \citenamefont {Buice}, \citenamefont {Casey}, \citenamefont {Chow},
  \citenamefont {Donahue}, \citenamefont {Ehrlichman}, \citenamefont {Harkins},
  \citenamefont {Hellert}, \citenamefont {Johnson}, \citenamefont {Jung},
  \citenamefont {Leemann}, \citenamefont {Leftwich-Vann}, \citenamefont
  {Leitner}, \citenamefont {Luo}, \citenamefont {Omolayo}, \citenamefont
  {Osborn}, \citenamefont {Penn}, \citenamefont {Portmann}, \citenamefont
  {Robin}, \citenamefont {Sannibale}, \citenamefont {Santis}, \citenamefont
  {Sun}, \citenamefont {Swenson}, \citenamefont {Venturini}, \citenamefont
  {Virostek}, \citenamefont {Waldron},\ and\ \citenamefont
  {Wallén}}]{alsu_steier_ipac2019}%
  \BibitemOpen
  \bibfield  {author} {\bibinfo {author} {\bibfnamefont {C.}~\bibnamefont
  {Steier}}, \bibinfo {author} {\bibfnamefont {P.}~\bibnamefont {Amstutz}},
  \bibinfo {author} {\bibfnamefont {K.}~\bibnamefont {Baptiste}}, \bibinfo
  {author} {\bibfnamefont {P.}~\bibnamefont {Bong}}, \bibinfo {author}
  {\bibfnamefont {E.}~\bibnamefont {Buice}}, \bibinfo {author} {\bibfnamefont
  {P.}~\bibnamefont {Casey}}, \bibinfo {author} {\bibfnamefont
  {K.}~\bibnamefont {Chow}}, \bibinfo {author} {\bibfnamefont {R.}~\bibnamefont
  {Donahue}}, \bibinfo {author} {\bibfnamefont {M.}~\bibnamefont {Ehrlichman}},
  \bibinfo {author} {\bibfnamefont {J.}~\bibnamefont {Harkins}}, \bibinfo
  {author} {\bibfnamefont {T.}~\bibnamefont {Hellert}}, \bibinfo {author}
  {\bibfnamefont {M.}~\bibnamefont {Johnson}}, \bibinfo {author} {\bibfnamefont
  {J.-Y.}\ \bibnamefont {Jung}}, \bibinfo {author} {\bibfnamefont
  {S.}~\bibnamefont {Leemann}}, \bibinfo {author} {\bibfnamefont
  {R.}~\bibnamefont {Leftwich-Vann}}, \bibinfo {author} {\bibfnamefont
  {D.}~\bibnamefont {Leitner}}, \bibinfo {author} {\bibfnamefont
  {T.}~\bibnamefont {Luo}}, \bibinfo {author} {\bibfnamefont {O.}~\bibnamefont
  {Omolayo}}, \bibinfo {author} {\bibfnamefont {J.}~\bibnamefont {Osborn}},
  \bibinfo {author} {\bibfnamefont {G.}~\bibnamefont {Penn}}, \bibinfo {author}
  {\bibfnamefont {G.}~\bibnamefont {Portmann}}, \bibinfo {author}
  {\bibfnamefont {D.}~\bibnamefont {Robin}}, \bibinfo {author} {\bibfnamefont
  {F.}~\bibnamefont {Sannibale}}, \bibinfo {author} {\bibfnamefont {S.~D.}\
  \bibnamefont {Santis}}, \bibinfo {author} {\bibfnamefont {C.}~\bibnamefont
  {Sun}}, \bibinfo {author} {\bibfnamefont {C.}~\bibnamefont {Swenson}},
  \bibinfo {author} {\bibfnamefont {M.}~\bibnamefont {Venturini}}, \bibinfo
  {author} {\bibfnamefont {S.}~\bibnamefont {Virostek}}, \bibinfo {author}
  {\bibfnamefont {W.}~\bibnamefont {Waldron}},\ and\ \bibinfo {author}
  {\bibfnamefont {E.}~\bibnamefont {Wallén}},\ }in\ \href
  {https://doi.org/10.18429/JACoW-IPAC2019-TUPGW097} {\emph {\bibinfo
  {booktitle} {Proc. IPAC'19}}}\ (\bibinfo  {publisher} {JACoW Publishing,
  Geneva, Switzerland},\ \bibinfo {year} {2019})\ pp.\ \bibinfo {pages}
  {1639--1642}\BibitemShut {NoStop}%
\bibitem [{\citenamefont {Hellert}(2026)}]{autoresearch_code}%
  \BibitemOpen
  \bibfield  {author} {\bibinfo {author} {\bibfnamefont {T.}~\bibnamefont
  {Hellert}},\ }\href {https://github.com/als-apg/autoresearch-commissioning}
  {\bibinfo {title} {\texttt{autoresearch}: autonomous discovery of accelerator
  commissioning algorithms}},\ \bibinfo {howpublished}
  {\url{https://github.com/als-apg/autoresearch-commissioning}} (\bibinfo
  {year} {2026}),\ \bibinfo {note} {research loop, simulated-commissioning
  harness, and campaign configurations}\BibitemShut {NoStop}%
\bibitem [{\citenamefont {Malina}\ \emph {et~al.}(2023)\citenamefont {Malina},
  \citenamefont {Agapov}, \citenamefont {Keil}, \citenamefont {Musa},
  \citenamefont {Veglia}, \citenamefont {Carmignani}, \citenamefont {Carver},
  \citenamefont {Hoummi}, \citenamefont {Liuzzo}, \citenamefont {Perron},
  \citenamefont {White},\ and\ \citenamefont {Hellert}}]{petraivesr2023_27}%
  \BibitemOpen
  \bibfield  {author} {\bibinfo {author} {\bibfnamefont {L.}~\bibnamefont
  {Malina}}, \bibinfo {author} {\bibfnamefont {I.}~\bibnamefont {Agapov}},
  \bibinfo {author} {\bibfnamefont {J.}~\bibnamefont {Keil}}, \bibinfo {author}
  {\bibfnamefont {E.~S.~H.}\ \bibnamefont {Musa}}, \bibinfo {author}
  {\bibfnamefont {B.}~\bibnamefont {Veglia}}, \bibinfo {author} {\bibfnamefont
  {N.}~\bibnamefont {Carmignani}}, \bibinfo {author} {\bibfnamefont
  {L.}~\bibnamefont {Carver}}, \bibinfo {author} {\bibfnamefont
  {L.}~\bibnamefont {Hoummi}}, \bibinfo {author} {\bibfnamefont {S.~M.}\
  \bibnamefont {Liuzzo}}, \bibinfo {author} {\bibfnamefont {T.}~\bibnamefont
  {Perron}}, \bibinfo {author} {\bibfnamefont {S.}~\bibnamefont {White}},\ and\
  \bibinfo {author} {\bibfnamefont {T.}~\bibnamefont {Hellert}},\ }in\ \href
  {https://doi.org/10.18429/JACoW-ICALEPCS2023-FR2AO05} {\emph {\bibinfo
  {booktitle} {Proc. ICALEPCS'23}}}\ (\bibinfo  {publisher} {JACoW Publishing,
  Geneva, Switzerland},\ \bibinfo {year} {2023})\ pp.\ \bibinfo {pages}
  {1637--1642}\BibitemShut {NoStop}%
\bibitem [{\citenamefont {Liuzzo}\ \emph {et~al.}(2024)\citenamefont {Liuzzo},
  \citenamefont {Carmignani}, \citenamefont {Carver}, \citenamefont {Hoummi},
  \citenamefont {Perron}, \citenamefont {White}, \citenamefont {Agapov},
  \citenamefont {Boese}, \citenamefont {Hellert}, \citenamefont {Keil},
  \citenamefont {Malina}, \citenamefont {Musa},\ and\ \citenamefont
  {Veglia}}]{esrfebspet2024_16}%
  \BibitemOpen
  \bibfield  {author} {\bibinfo {author} {\bibfnamefont {S.~M.}\ \bibnamefont
  {Liuzzo}}, \bibinfo {author} {\bibfnamefont {N.}~\bibnamefont {Carmignani}},
  \bibinfo {author} {\bibfnamefont {L.~R.}\ \bibnamefont {Carver}}, \bibinfo
  {author} {\bibfnamefont {L.}~\bibnamefont {Hoummi}}, \bibinfo {author}
  {\bibfnamefont {T.}~\bibnamefont {Perron}}, \bibinfo {author} {\bibfnamefont
  {S.}~\bibnamefont {White}}, \bibinfo {author} {\bibfnamefont
  {I.}~\bibnamefont {Agapov}}, \bibinfo {author} {\bibfnamefont
  {M.}~\bibnamefont {Boese}}, \bibinfo {author} {\bibfnamefont
  {T.}~\bibnamefont {Hellert}}, \bibinfo {author} {\bibfnamefont
  {J.}~\bibnamefont {Keil}}, \bibinfo {author} {\bibfnamefont {L.}~\bibnamefont
  {Malina}}, \bibinfo {author} {\bibfnamefont {E.}~\bibnamefont {Musa}},\ and\
  \bibinfo {author} {\bibfnamefont {B.}~\bibnamefont {Veglia}},\ }\href
  {https://doi.org/10.1088/1742-6596/2687/3/032001} {\bibfield  {journal}
  {\bibinfo  {journal} {J. Phys. Conf. Ser.}\ }\textbf {\bibinfo {volume}
  {2687}},\ \bibinfo {pages} {032001} (\bibinfo {year} {2024})}\BibitemShut
  {NoStop}%
\bibitem [{\citenamefont {Di~Mitri}(2023)}]{dimitri_fundamentals_2023}%
  \BibitemOpen
  \bibfield  {author} {\bibinfo {author} {\bibfnamefont {S.}~\bibnamefont
  {Di~Mitri}},\ }\href {https://doi.org/10.1007/978-3-031-07662-6} {\emph
  {\bibinfo {title} {{Fundamentals of Particle Accelerator Physics}}}},\
  Graduate Texts in Physics\ (\bibinfo  {publisher} {Springer Cham},\ \bibinfo
  {year} {2023})\BibitemShut {NoStop}%
\bibitem [{\citenamefont {Wiedemann}(2015)}]{wiedemann_pap_2015}%
  \BibitemOpen
  \bibfield  {author} {\bibinfo {author} {\bibfnamefont {H.}~\bibnamefont
  {Wiedemann}},\ }\href {https://doi.org/10.1007/978-3-319-18317-6} {\emph
  {\bibinfo {title} {{Particle Accelerator Physics}}}},\ \bibinfo {edition}
  {4th}\ ed.,\ Graduate Texts in Physics\ (\bibinfo  {publisher} {Springer
  Cham},\ \bibinfo {year} {2015})\BibitemShut {NoStop}%
\bibitem [{\citenamefont {{Anthropic}}(2025)}]{anthropic-haiku45}%
  \BibitemOpen
  \bibfield  {author} {\bibinfo {author} {\bibnamefont {{Anthropic}}},\ }\href
  {https://assets.anthropic.com/m/99128ddd009bdcb/original/Claude-Haiku-4-5-System-Card.pdf}
  {\bibinfo {title} {System card: {Claude} {Haiku} 4.5}},\ \bibinfo
  {howpublished}
  {\url{https://assets.anthropic.com/m/99128ddd009bdcb/original/Claude-Haiku-4-5-System-Card.pdf}}
  (\bibinfo {year} {2025})\BibitemShut {NoStop}%
\bibitem [{\citenamefont
  {{Anthropic}}(2026{\natexlab{a}})}]{anthropic-sonnet46}%
  \BibitemOpen
  \bibfield  {author} {\bibinfo {author} {\bibnamefont {{Anthropic}}},\ }\href
  {https://www.anthropic.com/research/claude-sonnet-4-6} {\bibinfo {title}
  {System card: {Claude} {Sonnet} 4.6}},\ \bibinfo {howpublished}
  {\url{https://www.anthropic.com/research/claude-sonnet-4-6}} (\bibinfo {year}
  {2026}{\natexlab{a}})\BibitemShut {NoStop}%
\bibitem [{\citenamefont {{Anthropic}}(2026{\natexlab{b}})}]{anthropic-opus46}%
  \BibitemOpen
  \bibfield  {author} {\bibinfo {author} {\bibnamefont {{Anthropic}}},\ }\href
  {https://www.anthropic.com/news/claude-opus-4-6} {\bibinfo {title} {System
  card: {Claude} {Opus} 4.6}},\ \bibinfo {howpublished}
  {\url{https://www.anthropic.com/news/claude-opus-4-6}} (\bibinfo {year}
  {2026}{\natexlab{b}})\BibitemShut {NoStop}%
\bibitem [{\citenamefont {Deb}(2001)}]{deb_moo_2001}%
  \BibitemOpen
  \bibfield  {author} {\bibinfo {author} {\bibfnamefont {K.}~\bibnamefont
  {Deb}},\ }\href@noop {} {\emph {\bibinfo {title} {{Multi-Objective
  Optimization Using Evolutionary Algorithms}}}}\ (\bibinfo  {publisher} {John
  Wiley \& Sons},\ \bibinfo {address} {Chichester},\ \bibinfo {year}
  {2001})\BibitemShut {NoStop}%
\bibitem [{\citenamefont {Amodei}\ \emph {et~al.}(2016)\citenamefont {Amodei},
  \citenamefont {Olah}, \citenamefont {Steinhardt}, \citenamefont {Christiano},
  \citenamefont {Schulman},\ and\ \citenamefont
  {Man{\'e}}}]{amodei_concrete_2016}%
  \BibitemOpen
  \bibfield  {author} {\bibinfo {author} {\bibfnamefont {D.}~\bibnamefont
  {Amodei}}, \bibinfo {author} {\bibfnamefont {C.}~\bibnamefont {Olah}},
  \bibinfo {author} {\bibfnamefont {J.}~\bibnamefont {Steinhardt}}, \bibinfo
  {author} {\bibfnamefont {P.}~\bibnamefont {Christiano}}, \bibinfo {author}
  {\bibfnamefont {J.}~\bibnamefont {Schulman}},\ and\ \bibinfo {author}
  {\bibfnamefont {D.}~\bibnamefont {Man{\'e}}},\ }\href@noop {} {\bibfield
  {journal} {\bibinfo  {journal} {arXiv preprint arXiv:1606.06565}\ } (\bibinfo
  {year} {2016})},\ \Eprint {https://arxiv.org/abs/1606.06565}
  {arXiv:1606.06565 [cs.AI]} \BibitemShut {NoStop}%
\bibitem [{\citenamefont {Leike}\ \emph {et~al.}(2017)\citenamefont {Leike},
  \citenamefont {Martic}, \citenamefont {Krakovna}, \citenamefont {Ortega},
  \citenamefont {Everitt}, \citenamefont {Lefrancq}, \citenamefont {Orseau},\
  and\ \citenamefont {Legg}}]{leike_gridworlds_2017}%
  \BibitemOpen
  \bibfield  {author} {\bibinfo {author} {\bibfnamefont {J.}~\bibnamefont
  {Leike}}, \bibinfo {author} {\bibfnamefont {M.}~\bibnamefont {Martic}},
  \bibinfo {author} {\bibfnamefont {V.}~\bibnamefont {Krakovna}}, \bibinfo
  {author} {\bibfnamefont {P.~A.}\ \bibnamefont {Ortega}}, \bibinfo {author}
  {\bibfnamefont {T.}~\bibnamefont {Everitt}}, \bibinfo {author} {\bibfnamefont
  {A.}~\bibnamefont {Lefrancq}}, \bibinfo {author} {\bibfnamefont
  {L.}~\bibnamefont {Orseau}},\ and\ \bibinfo {author} {\bibfnamefont
  {S.}~\bibnamefont {Legg}},\ }\href@noop {} {\bibfield  {journal} {\bibinfo
  {journal} {arXiv preprint arXiv:1711.09883}\ } (\bibinfo {year} {2017})},\
  \Eprint {https://arxiv.org/abs/1711.09883} {arXiv:1711.09883 [cs.LG]}
  \BibitemShut {NoStop}%
\bibitem [{\citenamefont {Skalse}\ \emph {et~al.}(2022)\citenamefont {Skalse},
  \citenamefont {Howe}, \citenamefont {Krasheninnikov},\ and\ \citenamefont
  {Krueger}}]{skalse_reward_2022}%
  \BibitemOpen
  \bibfield  {author} {\bibinfo {author} {\bibfnamefont {J.}~\bibnamefont
  {Skalse}}, \bibinfo {author} {\bibfnamefont {N.~H.~R.}\ \bibnamefont {Howe}},
  \bibinfo {author} {\bibfnamefont {D.}~\bibnamefont {Krasheninnikov}},\ and\
  \bibinfo {author} {\bibfnamefont {D.}~\bibnamefont {Krueger}},\ }\href@noop
  {} {\bibfield  {journal} {\bibinfo  {journal} {Advances in Neural Information
  Processing Systems}\ }\textbf {\bibinfo {volume} {35}},\ \bibinfo {pages}
  {9460} (\bibinfo {year} {2022})}\BibitemShut {NoStop}%
\end{thebibliography}%

\end{document}